\documentstyle[eqsecnum,prd,aps,epsf]{revtex}

\draft
\begin{document}

\def\beqa{\begin{eqnarray}}
\def\eeqa{\end{eqnarray}}
\def\bI{\hbox{$\,I\!\!\!\!-$}}
\def\a{\alpha}
\def\b{\beta}
\def\p{\partial}
\def\e{\epsilon}
\def\ve{\varepsilon}
\def\r{\rho}
\def\O{\Omega}
\def\t{\tilde}
\def\ra{\rightarrow}
\newcommand{\beq}{\begin{equation}}
\newcommand{\eeq}{\end{equation}}
\newcommand{\beqn}{\begin{eqnarray}}
\newcommand{\eeqn}{\end{eqnarray}}
\newcommand{\pa}{\partial}

\thispagestyle{empty}
{\baselineskip0pt
\leftline{\large\baselineskip16pt\sl\vbox to0pt{\hbox{\it Department of Physics}
               \hbox{\it Kyoto University}\vss}}
\rightline{\baselineskip16pt\rm\vbox to20pt{\hbox{KUNS 1464}
           \hbox{OU-TAP 68}
           \hbox{July, 1998}
\vss}}%
}
\vskip1cm
\begin{center}{\large \bf
Gravitational radiation from corotating binary neutron stars of
incompressible fluid in the first post-Newtonian approximation of
general relativity
}
\end{center}

\vskip 0.4cm

\begin{center}
Keisuke Taniguchi 
\footnote{Electronic address: taniguci@tap.scphys.kyoto-u.ac.jp}~
and ~
Masaru Shibata 
\footnote{Electronic address: shibata@vega.ess.sci.osaka-u.ac.jp}
\end{center} 

\begin{center}
$^*${\em Department of Physics,~Kyoto University,~Kyoto 606-8502,~Japan} \\
$^\dagger$
{\em Department of Earth and Space Science,~Graduate School of
  Science,~Osaka University,~Toyonaka,~Osaka 560-0043,~Japan}
\end{center}

\begin{abstract}

  We analytically study gravitational radiation from corotating
  binary neutron stars composed of incompressible, homogeneous fluid in
  circular orbits. The energy and the angular momentum loss rates are
  derived up to the first post-Newtonian (1PN) 
  order beyond the quadrupole approximation
  including effects of the finite size of each star of binary. It is found
  that the leading term of finite size effects in the 1PN order is only
  $O(GM_{\ast}/c^2 a_{\ast})$ smaller than that in the Newtonian order,
  where $GM_{\ast}/c^2 a_{\ast}$ means the ratio of the gravitational
  radius to the mean radius of each star of binary, and the 1PN term acts 
  to decrease the Newtonian finite size effect in gravitational radiation. 
\end{abstract}


\section{Introduction}

Around the beginning of the next century, the laser interferometers for
detection of gravitational waves, such as LIGO\cite{LIGO},
VIRGO\cite{VIRGO}, GEO600\cite{GEO} and TAMA300\cite{TAMA} will be 
in operation. The most important targets for these
detectors are coalescing binary neutron stars (BNS's) because in a 
period of their 
inspiraling phase they will emit gravitational waves of frequencies in
the sensitive range of these detectors, i.e., from 10Hz to 1000Hz. We
have to prepare accurate theoretical templates in order to extract
informations of BNS's such as each mass, spin and so on using
matched filtering technique\cite{CF}.

In their early inspiraling phase, the hydrodynamical effect of each star
of binary is less effective and the point particle approximation works
well. In the point particle approximation, the energy
loss rate is calculated up to 2.5 post-Newtonian (PN) order for
arbitrary mass binaries\cite{BDIWW}. The other approach for the
inspiraling phase is the black hole perturbation method. In this
formalism, the energy loss rate is calculated up to 4PN order in a Kerr
black hole case\cite{TSTS}, and 5.5PN order in a Schwarzschild black
hole case\cite{TTS}.

In the late inspiraling phase, however, the hydrodynamical effects of
each star of binary become important, and we have to study the
evolution of the binary taking into account them\cite{LRS}
\cite{TS}\cite{Lombardi}\cite{Shibata}.  The importance of the study of
the late inspiraling phase comes from the existence of the innermost
stable circular orbit (ISCO). If we are able to 
know the location of the ISCO from the signal of gravitational waves, 
we can see a strong general relativistic phenomenon as well as we 
may get the information of the equation of state of neutron stars, i.e.,
the relation between the mass and the radius\cite{Lindblom}\cite{ZCM}.
To know the precise location of the ISCO, 
it is desirable to construct the theoretical
template of gravitational waves 
in the late inspiraling phase as in the early one. Up to the
Newtonian order, we have already understood gravitational radiation from
binary systems of finite size 
stars\cite{CC}\cite{Nfinite}\cite{BC}, but apparently the Newtonian 
treatment is not appropriate because BNS's are general relativistic 
objects. 

To investigate general relativistic effects on 
gravitational waves in the late 
inspiraling phase, in this paper, we analytically calculate the
energy and angular momentum loss rates of the 1PN order including the
finite size effects. In the calculation, we assume that BNS's are 
composed of the incompressible and homogeneous fluid 
and then obtain an ellipsoidal equilibrium configuration for each star. 
Also, we suppose the corotating circular orbit in which each star
of binary uniformly rotates around the center of mass of the system. 
Although the
assumption of corotation is not appropriate for the realistic one which
will be almost irrotational\cite{Kochanek}\cite{BC}, the energy and angular
momentum loss rates we calculate in this paper will be useful to see 
contribution of the finite size effects.

This paper is organized as follows. In Sec. II, we derive the mass
quadrupole moments and their time derivatives which 
are necessary for calculation of gravitational waves by 
using Blanchet-Damour formalism\cite{Blanchet}. In Sec. III, we give the
rates of the energy and angular momentum loss gathering the time
derivatives of the mass quadrupole moments which we derive in Sec. II,
and their quantitative results are given in Sec. IV. Section V is
devoted to summary and discussion.

Throughout this paper, we use the unit $G=1$ and $c$ denotes the speed
of light. Latin indices $i,~j,~k,\cdots$ take 1 to 3, and $\delta_{ij}$
denotes the Kronecker's delta. We use
$I_{ij}$ and $\bI_{ij}$ as the quadrupole moment and its trace free
part,
\beqa
  I_{ij} &\equiv& \int d^3 x \r x_i x_j ={M \over 5} a_i^2 \delta_{ij}, \\
  \bI_{ij} &=& I_{ij} -{1 \over 3} \delta_{ij} \sum_{k=1}^3 I_{kk},
\eeqa
where $M$ and $a_i$ denote the Newtonian mass and 
the principal axes of the star. 

\section{Formulation} \label{formulation}

In this paper, we calculate the energy and angular momentum loss rates
of gravitational radiation from corotating binary stars. 
Each star of binary is assumed to be composed of
the incompressible and homogeneous fluid because all the calculations can
be done analytically. We pay particular attention to BNS's of equal
masses in circular orbits whose coordinate separation is $R$. We assume 
that the center of mass of the binary system locates at the origin of 
the {\it inertial} coordinate system $(X_1, X_2, X_3)$, 
and the center of mass of a 
star (star 1) locates at the origin of the {\it comoving} coordinate 
system $(x_1, x_2, x_3)$. 
In the comoving coordinate system, the
other star (star 2) locates at $(x_1, x_2, x_3)=(-R, 0, 0)$. The
transformation from the coordinate system $(X_1, X_2, X_3)$ to
$(x_1, x_2, x_3)$ is given by
\beqa
  X_1 &=& \biggl( x_1 +{R \over 2} \biggr) \cos \O t -x_2 \sin \O t,
  \nonumber \\
  X_2 &=& \biggl( x_1 +{R \over 2} \biggr) \sin \O t +x_2 \cos \O t,
  \nonumber \\
  X_3 &=& x_3. \label{transfer}
\eeqa

According to Blanchet and Damour\cite{Blanchet}, 
in the 1PN order, the energy and angular
momentum loss rates are written as
\beqa
  {dE \over dt} &=& -{1 \over 5c^5} \biggl[ M_{ij}^{(3)} M_{ij}^{(3)} +
  {1 \over c^2} \biggl( {5 \over 189} M_{ijk}^{(4)} M_{ijk}^{(4)} +
  {16 \over 9} S_{ij}^{(3)} S_{ij}^{(3)} \biggr) \biggr],
  \label{dedtdef} \\
  {dJ^3 \over dt} &=&- {2 \over 5 c^5} \epsilon_{3jk} \biggl[ M_{jl}^{(2)} 
  M_{kl}^{(3)} + {1 \over c^2} \biggl( {5 \over 126} M_{jlm}^{(3)}
  M_{klm}^{(4)} + {16 \over 9} S_{jl}^{(2)} S_{kl}^{(3)} \biggr)
  \biggr], \label{djdtdef}
\eeqa
where $M_{ij \cdots}$ and $S_{ij}$ are Blanchet-Damour's mass multipole 
moments and the current quadrupole moment, and the superscript 
$(n)$ where $n=2$, 3, 4 denotes the time derivative
$d^n/dt^n$. Since we assume that the binary 
rotates around the $X_3$ axis, 
there is only $i=3$ component in the angular momentum loss rate. 
In the subsequent subsections, we derive 
$M_{ij \cdots}$ and $S_{ij}$ which are necessary for calculation of
gravitational waves.

\subsection{Mass quadrupole moment}

First, we calculate $M_{ij}$ up to the 1PN order. $M_{ij}$ is the even
term of the inertial coordinates $X_1$, $X_2$, and $X_3$, 
and we consider the equal mass binary. Then, the contribution from star 1 
is the same as that from star 2. $M_{ij}$ is expressed as
\beqa
  M_{ij} (t) &=& \int d^3 X \sigma \hat{X}_{ij} + {1 \over 14c^2} {d^2
  \over dt^2} \biggl( \int d^3 X \r \hat{X}_{ij} {\bf X}^2 \biggr)
  -{20 \over 21 c^2} {d \over dt} \biggl( \int d^3 X \r v_k \hat{X}_{ijk} 
  \biggr), \nonumber \\
  &=& \int d^3 X \r \hat{X}_{ij}+{1 \over c^2} \bigg[ \int d^3 
    X \r \biggl( 2v^2 +2U +{3P \over \r} \biggr) \hat{X}_{ij}
  + {1 \over 14} {d^2 \over dt^2} \biggl( \int d^3 X \r \hat{X}_{ij} {\bf
      X}^2 \biggr) \nonumber \\
  & &\hspace{100pt} -{20 \over 21} {d \over dt}
    \biggl( \int d^3 X \r v_k \hat{X}_{ijk} \biggr) \biggr],
  \label{massquadmom}
\eeqa
where
\beqa
  \sigma &=& \r \biggl[ 1+ {1 \over c^2} \biggl( 2v^2 +2U +{3P \over \r}
  \biggr) \biggr], \\
  \hat{X}_{ij} &=& X_i X_j -{1 \over 3} \delta_{ij} {\bf X}^2, \\
  \hat{X}_{ijk} &=& X_i X_j X_k -{1 \over 5} {\bf X}^2 (\delta_{ij} X_k
  + \delta_{jk} X_i +\delta_{ki} X_j ), \\
  {\bf X}^2 &=& X_1^2 +X_2^2 +X_3^2, \\
  v^i &=& (-\O X_2,~ \O X_1,~ 0). 
\eeqa
In the following calculation, we integrate only over
star 1 because the total mass quadrupole moment is twice as 
large as that of star 1. Moreover, as shown in Ref. \cite{TS}, the
deformation from the Newtonian ellipsoid becomes higher order effect of
$R^{-1}$. Therefore, we integrate over the ellipsoid simply.

\subsubsection{Newtonian order}

The quadrupole moment of the Newtonian order appears in the first term of
Eq. (\ref{massquadmom}) and expressed as
\beqa
  D_{ij} \equiv \int d^3 X \r \biggl(X_i X_j -{1 \over 3} \delta_{ij}
  {\bf X}^2 \biggr).
\eeqa
Using Eq. (\ref{transfer}), we get the components as
\beqa
  D_{11} &=& \hat{D} \cos^2 \O t +{\rm const.}, \\
  D_{22} &=&-\hat{D} \sin^2 \O t +{\rm const.}, \\
  D_{12} &=& \hat{D} \sin \O t \cos \O t,
\eeqa
where $\hat D = I_{11}-I_{22}+R^2M/4$. In these equations, we use 
$M \equiv \int \r d^3 x$ as the Newtonian mass. 
Since $D_{33}$ does not depend on time and the other terms vanish, we do
not write them here. Note that in these components, only the 
contribution from star 1 is included.

\subsubsection{1PN order}

Next, we calculate the 1PN order terms.
For the 1PN order terms in Eq. (\ref{massquadmom}), we define
\beqa
  V_{ij} &\equiv& 2 \int d^3 X \r v^2 \hat{X}_{ij}, \\
  U_{ij}^{1 \ra 1} &\equiv& 2 \int d^3 X \r U^{1 \ra 1} \hat{X}_{ij}, \\
  U_{ij}^{2 \ra 1} &\equiv& 2 \int d^3 X \r U^{2 \ra 1} \hat{X}_{ij}, \\
  P_{ij} &\equiv& 3 \int d^3 X \r {P \over \r} \hat{X}_{ij}, \\
  Q_{ij} &\equiv& {1 \over 14} {d^2 \over dt^2} \biggl( \int d^3 X \r
    \hat{X}_{ij} {\bf X}^2 \biggr), \\
  R_{ij} &\equiv& -{20 \over 21} {d \over dt}
    \biggl( \int d^3 X \r v_k \hat{X}_{ijk} \biggr),
\eeqa
because it is convenient to calculate the terms separately. In these
equations, we use
\beqa
  U^{1 \ra 1} &=& \pi \r \Bigl( A_0 -\sum_l A_l x_l^2 \Bigr), \\
  U^{2 \ra 1} &=& {M \over R} \biggl[ 1 -{x_1 \over R} +{2x_1^2 -x_2^2
    -x_3^2 \over 2R^2} +{-2 x_1^3 +3x_1 (x_2^2 +x_3^2) \over 2R^3}
  +O(R^{-4}) \biggr] \nonumber \\
  & &+ {3 \bI_{11} \over 2R^3} \biggl( 1 -{3x_1 \over R} +O(R^{-2})
    \biggr) +O(R^{-5}), \\
  P &=& P_0 \biggl( 1- \sum_l {x_l^2 \over a_l^2} \biggr), \\
  P_0 &=& {\r \over 3} \biggl[ \pi \r A_0 -{\O_{\rm N}^2 \over 2} (a_1^2 
  +a_2^2) -{M \over 2R^3} (2a_1^2 -a_2^2 -a_3^2) \biggr] +O(R^{-5}),
\eeqa
where $U^{1 \ra 1}$, $U^{2 \ra 1}$ and $P$ denote the Newtonian 
potential generated by star 1 itself, the Newtonian potential 
generated by star 2, and the pressure,
respectively\cite{TS}. $A_{ij \cdots}$ are index symbols introduced by
Chandrasekhar\cite{CH69} and $A_0=\sum_l A_l a_l^2$ is 
calculated from\cite{CH69}
\beqa
  A_0 &=& a_1 a_2 a_3 \int_0^{\infty} {du \over \sqrt{(a_1^2 +u) (a_2^2
      +u) (a_3^2 +u)}}, \\
  &=& a_1^2 \a_2 \a_3 \int_0^{\infty} {dt \over \sqrt{(1+t) (\a_2^2 +t)
      (\a_3^2 +t)}} \equiv a_1^2 \t{A}_0,
\eeqa
where $\a_2=a_2/a_1$ and $\a_3=a_3/a_1$ are axial ratios.

When we derive the 1PN order terms, we need to calculate only some
combinations: 
The 1PN order terms in $M_{ij}^{(3)} M_{ij}^{(3)}$ are given by
\beqa
  & &D_{ij}^{(3)} \Bigl( V_{ij}^{(3)} + U_{ij}^{1 \ra 1(3)} + U_{ij}^{2 \ra
    1(3)} + P_{ij}^{(3)} + Q_{ij}^{(3)} + R_{ij}^{(3)} \Bigr) \nonumber \\
  &=& D_{11}^{(3)} \biggl[ V_{11}^{(3)}-V_{22}^{(3)} + U_{11}^{1 \ra
    1(3)}-U_{22}^{1 \ra 1(3)} + U_{11}^{2 \ra 1(3)}- U_{22}^{2 \ra 1(3)} 
  + P_{11}^{(3)}-P_{22}^{(3)} + Q_{11}^{(3)}-Q_{22}^{(3)} +
  R_{11}^{(3)}- R_{22}^{(3)} \biggr] \nonumber \\
  & &+ 2D_{12}^{(3)} \biggl[ V_{12}^{(3)} + U_{12}^{1
    \ra 1(3)} + U_{12}^{2 \ra 1(3)} + P_{12}^{(3)} + Q_{12}^{(3)} +
  R_{12}^{(3)} \biggr], \label{aaaa}
\eeqa
where we use the relation $D_{22}^{(i)} =-D_{11}^{(i)}$. Then,
we need to calculate only two combinations of the components expressed
as $[ (1,1)-(2,2) ]$ and $[ (1,2) ]$. We separately show the terms of 
the 1PN order in Eq. (\ref{aaaa}) in the following.

\begin{description}
\item[(1.1)] $V_{11}-V_{22}$:

\beqa
  V_{11}-V_{22} &=& 2\O^2 \int d^3 X \r (X_1^2+X_2^2) (X_1^2-X_2^2)
  \nonumber \\
  &=& 2\O^2  \hat{V} \cos 2\O t,
\eeqa
where
\beqa
  \hat{V} ={R^4 M \over 16} +{3 R^2 \over 2} I_{11}+I_{1111} -I_{2222},
\eeqa
and 
\beqa
  I_{ijkl} &\equiv &\int d^3 x \r x_i x_j x_k x_l, \nonumber \\
   &=&{ M \over 35}\times \left\{
  \begin{array}{ll}
  3 a_i^4  & (i=j=k=l) \\
  a_i^2 a_k^2 & (i=j \ne k=l).
  \end{array}
\right.
\eeqa

\item[(1.2)] $V_{12}$:

\beqa
  V_{12} &=& 2\O^2 \int d^3 X \r (X_1^2+X_2^2) X_1 X_2 \nonumber \\
  &=& \O^2 \hat{V} \sin 2\O t.
\eeqa

\item[(2.1)] $U_{11}^{1 \ra 1}-U_{22}^{1 \ra 1}$:

\beqa
  U_{11}^{1 \ra 1}-U_{22}^{1 \ra 1} &=& 2 \int d^3 X \r U^{1 \ra 1}
  (X_1^2 -X_2^2) \nonumber \\
  &=& 2 \hat{U}^{1 \ra 1} \cos 2\O t,
\eeqa
where
\beqa
  \hat{U}^{1 \ra 1} =\pi \r \biggl[ A_0 \biggl(I_{11}-I_{22} +{R^2 M
  \over 4} \biggr) -\sum_l A_l \biggl( I_{11ll} -I_{22ll} +{R^2 \over 4} 
  I_{ll} \biggr) \biggr]. 
\eeqa

\item[(2.2)] $U_{12}^{1 \ra 1}$:

\beqa
  U_{12}^{1 \ra 1} &=& 2 \int d^3 X \r U^{1 \ra 1} X_1 X_2 \nonumber \\
  &=& \hat{U}^{1 \ra 1} \sin 2\O t. 
\eeqa

\item[(3.1)] $U_{11}^{2 \ra 1}-U_{22}^{2 \ra 1}$:

\beqa
  U_{11}^{2 \ra 1} -U_{22}^{2 \ra 1}
  &=& 2 \int d^3 X \r U^{2 \ra 1} (X_1^2 -X_2^2) \nonumber \\
  &=& 2 \hat{U}^{2 \ra 1} \cos 2\O t +O(R^{-3}), 
\eeqa
where
\beqa
  \hat{U}^{2 \ra 1} ={M \over R} \biggl( {R^2 M \over 4} -I_{22} +{3
  \over 4} \bI_{11} \biggr).
\eeqa

\item[(3.2)] $U_{12}^{2 \ra 1}$:

\beqa
  U_{12}^{2 \ra 1}
  &=& 2 \int d^3 X \r U^{2 \ra 1} X_1 X_2 \nonumber \\
  &=& \hat{U}^{2 \ra 1} \sin 2\O t +O(R^{-3}). 
\eeqa

\item[(4.1)] $P_{11}- P_{22}$:

\beqa
  P_{11}- P_{22} &=& 3\int d^3 X \r {P \over \r} (X_1^2-X_2^2) \nonumber 
  \\
  &=& 3 \hat{P} \cos 2 \O t,
\eeqa
where
\beqa
  \hat{P} ={P_0 \over \r} \biggl[ I_{11}-I_{22} +{R^2 M \over 4} -\sum_l
  {1 \over a_l^2} \biggl( I_{11ll} -I_{22ll} +{R^2 \over 4} I_{ll}
  \biggr) \biggr]. 
\eeqa

\item[(4.2)] $P_{12}$:

\beqa
  P_{12} &=& 3\int d^3 X \r {P \over \r} X_1 X_2 \nonumber \\
  &=& {3 \over 2} \hat{P} \sin 2 \O t. 
\eeqa

\item[(5.1)] $Q_{11}- Q_{22}$:

\beqa
  Q_{11}-Q_{22}
  &=& {1 \over 14} {d^2 \over dt^2} \biggl( \int d^3 X \r
  (X_1^2-X_2^2) {\bf X}^2 \biggr) \nonumber \\
  &=& -{2 \over 7} \O^2 \hat{Q} \cos 2\O t,
\eeqa
where
\beqa
  \hat{Q} = {R^4 M \over 16} +{R^2 \over
  4} (6I_{11}+I_{33})+I_{1111}-I_{2222}+I_{1133}-I_{2233}. 
\eeqa

\item[(5.2)] $Q_{12}$:

\beqa
  Q_{12}
  &=& {1 \over 14} {d^2 \over dt^2} \biggl( \int d^3 X \r
  X_1 X_2 {\bf X}^2 \biggr) \nonumber \\
  &=& -{1 \over 7} \O^2 \hat{Q} \sin 2\O t. 
\eeqa

\item[(6.1)] $R_{11}- R_{22}$:

\beqa
  R_{11}-R_{22}
  &=& -{16 \over 21} \O {d \over dt} \biggl( \int d^3 X \r
  X_1 X_2 {\bf X}^2 \biggr) \nonumber \\
  &=& -{16 \over 21} \O^2 \hat{Q} \cos 2\O t. 
\eeqa

\item[(6.2)] $R_{12}$:

\beqa
  R_{12} &=& {4 \over 21} \O {d \over dt} \biggl( \int d^3 X \r
  (X_1^2-X_2^2) {\bf X}^2 \biggr) \nonumber \\
  &=& -{8 \over 21} \O^2 \hat{Q} \sin 2\O t. 
\eeqa

\end{description}

\subsection{Mass octupole moment and current quadrupole
  moment}\label{momcqm}

In this subsection, we discuss on the mass octupole moment $M_{ijk}$ and
the current quadrupole one $S_{ij}$. They are written as
\beqa
  M_{ijk} (t) &=& \int d^3 X \r \hat{X}_{ijk}, \\
  S_{ij} (t) &=& \int d^3 X \r \e_{k l < i} \hat{X}_{j > k} v_l,
\eeqa
where
\beqa
  A_{<i} B_{j>} &=& {1 \over 2} (A_i B_j +B_i A_j ) -{1 \over 3}
  \delta_{ij} (A_k B_k).
\eeqa
$M_{ijk}$ and $S_{ij}$ are the odd terms of the coordinates $X_1$, $X_2$,
and $X_3$.  Therefore, the contribution from star 2 has the opposite
sign of that of star 1. Then, the total $M_{ijk}$ and $S_{ij}$ vanish as
\beqa
  (M_{ijk})_{\rm tot} &=& {}^1 M_{ijk} +{}^2 M_{ijk} =0, \\
  (S_{ij})_{\rm tot} &=& {}^1 S_{ij} +{}^2 S_{ij} =0,
\eeqa
where in the right hand side, superscripts 1 and 2 denote the 
contributions from star 1 and star 2, respectively. Therefore, they have no
contributions to the energy and the angular momentum loss rates 
in the case of the identical star binary.

\section{The energy and angular momentum loss rates}

Our purpose is to know the finite size effects of the 1PN order in
gravitational radiation. In the following, we calculate the energy 
and angular momentum loss rates up to $O(R^{-3})$ in the 1PN
order beyond the Newtonian order of the quadrupole formula. 

\subsection{The energy loss rate} \label{subsecelr}


Taking into account the fact stated in subsection \ref{momcqm}, the
energy loss rate is expressed as
\beqa
  {dE \over dt} &=&-{1 \over 5c^5} \biggl[ (M_{ij}^{(3)})_{\rm tot}
  (M_{ij}^{(3)})_{\rm tot} \biggr] \nonumber \\
  &=&-{1 \over 5c^5} \biggl[ (2D_{ij}^{(3)})
  (2D_{ij}^{(3)}) +{2 \over c^2} (2D_{ij}^{(3)}) \biggl\{ 2V_{ij}^{(3)}+
  2U_{ij}^{1 \ra 1(3)} + 2U_{ij}^{2 \ra 1(3)} + 2P_{ij}^{(3)} +
  2Q_{ij}^{(3)} + 2R_{ij}^{(3)} \biggr\} \biggr].
\eeqa

The contribution from the Newtonian order is
\beqa
  (2D_{ij}^{(3)}) (2D_{ij}^{(3)}) =128 \O^6 \hat{D}^2.
\eeqa

The contributions from the 1PN order are written as follows:
\beqa
  8 D_{ij}^{(3)} V_{ij}^{(3)} &=& 512 \O^8 \hat{D} \hat{V}, \\
  8 D_{ij}^{(3)} U_{ij}^{1 \ra 1 (3)}
  &=& 512 \O^6 \hat{D} \hat{U}^{1 \ra 1}, \\
  8 D_{ij}^{(3)} U_{ij}^{2 \ra 1 (3)}
  &=& 512 \O^6 \hat{D} \hat{U}^{2 \ra 1}, \\
  8 D_{ij}^{(3)} P_{ij}^{(3)} &=& 768 \O^6 \hat{D} \hat{P}, \\
  8 D_{ij}^{(3)} Q_{ij}^{(3)} &=& -{512 \over 7} \O^8 \hat{D} \hat{Q},
  \\
  8 D_{ij}^{(3)} R_{ij}^{(3)} &=& -{4096 \over 21} \O^8 \hat{D}
  \hat{Q}.
\eeqa

Collecting the results we derived above, we can get the energy loss rate
as
\beqa
  {dE \over dt} &=& -{64 M^5 \over 5 c^5 R^5} \Biggl[ 1+ {1 \over 5 R^2}
  \biggl\{ 8(a_1^2 - a_2^2) +9(2a_1^2 -a_2^2 -a_3^2) \biggr\} +O(R^{-4})
  \nonumber \\
  & &+{1 \over c^2} \biggl[ 10 \pi \rho A_0 -{151 M \over 84 R} + {4 \pi
    \rho A_0 \over 5 R^2} \biggl\{ {412 \over 21} (a_1^2-a_2^2) +{111
    \over 5} (2a_1^2 -a_2^2 -a_3^2) \biggr\} \nonumber \\
  & &\hspace{30pt}-{M \over 210 R^3}
  (1658a_1^2 -1103a_2^2 -16a_3^2) +O(R^{-4}) \biggr]
  \Biggr], \label{energylossrate}
\eeqa
where we use the angular velocity of the 1PN order\cite{TS}
\beqa
  \O^2 &=& {2M \over R^3} \biggl[ 1 +{1 \over c^2} \biggl\{ 2 \pi \r A_0 
  -{9M \over 4R} -{M \over 10R^3} (28a_1^2 -14a_2^2 -9a_3^2) +O(R^{-4})
  \biggr\} \biggr] \nonumber \\
  & &+{18 \bI_{11} \over R^5} \biggl( 1+ {28 \over 15c^2}
  \pi \r A_0 +O(R^{-2}) \biggr). \label{1PNomega}
\eeqa

To express Eq. (\ref{energylossrate}), we have used $M$, 
but it does not conserve through the sequence of the binary. 
Moreover, we simply used $R$ as the orbital separation, 
but in the 1PN case, the center of mass deviates from that 
in the Newtonian one and as a result, the 
orbital separation also deviates from $R$. 
We need a conserved mass and an appropriate 
definition of the center of mass. 

If we use the conserved {\it proper} mass 
\beqa
  M_{\ast} &=& \int d^3 x \r \biggl[ 1+ {1 \over c^2} \biggl( {v^2 \over
    2} +3U \biggr) \biggr] \nonumber \\
  &=& M \biggl[ 1+ {1 \over c^2} \biggl\{ {12 \pi \rho A_0 \over 5}
  +{13M \over 4R} +{M \over 20R^3} (34a_1^2 -11a_2^2 -15a_3^2)
  +O(R^{-5}) \biggr\} \biggr],
\eeqa
and define the center of mass by it as
\beqa
  x_{\ast}^i ={ 1 \over M_{\ast}} \int d^3 x \r_{\ast} x^i,
\eeqa
where
\beqa
  \r_{\ast} \equiv \r \biggl[ 1+ {1 \over c^2} \biggl( {v^2 \over
    2} +3U \biggr) \biggr],
\eeqa
the orbital separation should be replaced by
\beqa
  R_{\ast}=R \biggl[ 1+ {1 \over c^2} \biggl\{ -{4 M a_1^2 \over
    5R^3} +O(R^{-5}) \biggr\} \biggr].
\eeqa
Using these expressions, the energy loss rate is rewritten as
\beqa
  {dE \over dt} &=& -{64 M_{\ast}^5 \over 5 c^5 R_{\ast}^5} \Biggl[ 1+
  {1 \over 5 R_{\ast}^2} \biggl\{ 8(a_{1 \ast}^2 - a_{2 \ast}^2)
  +9(2a_{1 \ast}^2 -a_{2 \ast}^2 -a_{3 \ast}^2)
  \biggr\} +O(R_{\ast}^{-4}) \nonumber \\
  & &+{1 \over c^2} \biggl[ -2 \pi \rho A_{0\ast} -{379 M_{\ast} \over 21
    R_{\ast}} -{8 \pi \rho
    A_{0 \ast} \over 25 R_{\ast}^2} \biggl\{ {398 \over 21} (a_{1
    \ast}^2-a_{2 \ast}^2) +21
  (2a_{1 \ast}^2 -a_{2 \ast}^2 -a_{3 \ast}^2) \biggr\} \nonumber \\
  & &\hspace{30pt} -{M_{\ast} \over 210 R_{\ast}^3} (24394a_{1 \ast}^2
  -14830a_{2 \ast}^2 -7765a_{3 \ast}^2)
  +O(R_{\ast}^{-4}) \biggr] \Biggr], \label{dedt}
\eeqa
where we use the relation
\beqa
  a_i^2 =a_{i \ast}^2 \biggl[ 1 -{1 \over c^2} \biggl\{ {8 \pi \r A_{0
      \ast} \over 5} +{13M_{\ast} \over 6R} +{M_{\ast} \over 30R^3}
  (34a_{1 \ast}^2 -11a_{2 \ast}^2 -15a_{3 \ast}^2)
  \biggr\} \biggr], \label{aiastrelation}
\eeqa
and $a_{1\ast}a_{2\ast}a_{3\ast}=M_*/(4\pi\rho/3)=$ constant in the 1PN 
order. The expression of Eq. (\ref{dedt}) seems to depend
on the internal structure of the star of
binary even if we take the limits $a_i/R \rightarrow 0$. 
Thus, $M_*$ is not appropriate in comparing the 
energy luminosity with that for the point particle case. 
This is simply because $M_*$ is not the gravitational mass. 
Instead of $M_*$, we need an appropriate {\it gravitational} mass. 

If we use the parameterized post-Newtonian (PPN) mass\cite{Will} 
\beqa
  M_{{\rm PPN}} &=& \int d^3 x \rho \biggl[ 1 +{1 \over c^2} \biggl(
  {v^2 \over 2} +3U -{1 \over 2} U_{{\rm self}} +{v_{{\rm self}}^2
    \over 2} \biggr) \biggl] \nonumber \\
  &=& M \biggl[ 1 +{1 \over c^2} \biggl\{ 2 \pi \rho A_0 +{13M \over 4R}
  +{M \over 20 R^3} (38a_1^2 -7a_2^2 -15a_3^2) +O(R^{-5}) \biggr\}
  \biggr],
\eeqa
with the center of mass of each star defined as
\beqa
  x_{\rm PPN}^i ={1 \over M_{\rm PPN}} \int d^3 x \r x^i \biggl[ 1 +{1
    \over c^2} \biggl( {v^2 \over 2} +3U -{1 \over 2} U_{\rm self}
  +{v_{\rm self}^2 \over 2} \biggr) \biggr]
\eeqa
and the orbital separation as 
\beqa
  R_{\rm PPN} =R \biggl[ 1+ {1 \over c^2} \biggl\{ -{4 M a_1^2 \over
    5R^3} +O(R^{-5}) \biggr\} \biggr],
\eeqa
the leading internal structure dependent term in the 1PN order is
renormalized. Substituting the PPN mass and $R_{\rm PPN}$ into
Eq. (\ref{energylossrate}), we rewrite the energy loss rate as 
\beqa
  {dE \over dt} &=&-{64 M_{{\rm PPN}}^5 \over 5 c^5 R_{\rm PPN}^5}
  \Biggl[ 1+ {1 \over 5 R_{\rm PPN}^2} \biggl\{ 8(a_{1 {\rm PPN}}^2 -
  a_{2 {\rm PPN}}^2) +9(2a_{1 {\rm PPN}}^2 -a_{2 {\rm PPN}}^2 -a_{3 {\rm
      PPN}}^2) \biggr\} +O(R_{\rm PPN}^{-4}) \nonumber \\
  & &+{1 \over c^2} \biggl[ -{379 M_{{\rm PPN}} \over 21 R_{\rm PPN}}
  -{2 \pi \rho A_{0 {\rm PPN}} \over 15 R_{\rm PPN}^2} \biggl\{ {128
    \over 7} (a_{1 {\rm PPN}}^2-a_{2 {\rm PPN}}^2) +{99 \over 5}
  (2a_{1 {\rm PPN}}^2 -a_{2 {\rm PPN}}^2 -a_{3 {\rm PPN}}^2) \biggr\}
  \nonumber \\
  & &\hspace{30pt} -{M_{{\rm PPN}} \over 210 R_{\rm PPN}^3} (24604a_{1
    {\rm PPN}}^2 -14620a_{2 {\rm PPN}}^2 -7765a_{3 {\rm PPN}}^2)
  +O(R_{\rm PPN}^{-4}) \biggr] \Biggr], \label{dedtppn}
\eeqa
where we use the relation
\beqa
  a_i^2 =a_{i {\rm PPN}}^2 \biggl[ 1 -{1 \over c^2} \biggl\{ {4 \pi \r
    A_{0 {\rm PPN}}
    \over 3} +{13M_{\rm PPN} \over 6R} +{M_{\rm PPN} \over 30R^3}
  (38a_{1 {\rm PPN}}^2 -7a_{2 {\rm PPN}}^2 -15a_{3 {\rm PPN}}^2)
  \biggr\} \biggr]. \label{aippnrelation}
\eeqa
Then we recover the energy loss rate given for the point equal mass
binary case\cite{BDIWW} when we take the limits $a_i/R \rightarrow 0$.
(However, $M_{\rm PPN}$ appears to be the imperfect definition of 
gravitational mass as shown below.) 

{}From Eq. (\ref{dedtppn}), we find that the leading finite size effect
of 1PN order is only of $O(\rho A_0)(\sim O(M/a_i))$ 
smaller than the Newtonian finite size term. The next order terms are  
of $O(M/R)$ smaller than the Newtonian term. Here, we note that 
the effect of the spin-orbit (SO) coupling appears in
the same order as the latter terms for the case of the corotating binary
case. Including the SO coupling term, the energy loss rate of the
point particle binary of equal mass is written as\cite{Kidder}
\beqa
  {dE \over dt} =-{64 M_{\rm PPN}^5 \over 5 c^5 R_{\rm PPN}^5} \biggl[ 1
  -{1 \over c^2} \biggl\{ {379M_{\rm PPN} \over 21 R_{\rm PPN}} +{74
    M_{\rm PPN} \over 15 R_{\rm PPN}^3} (a_{1 {\rm PPN}}^2 +a_{2 {\rm
      PPN}}^2) \biggr\} \biggr].
\eeqa
We can see from this equation that the latter finite size terms 
of the 1PN order
in Eq. (\ref{dedtppn}) is not explained only by the SO coupling term. In
order to examine whether this fact is really true or not, we rewrite
$dE/dt$ using the angular velocity which is the invariant value of the
coordinate. 

Using the angular velocity, the orbital separation is written as
\beqa
    R &=& \biggl( {2 M_{\rm PPN} \over \O^2} \biggr)^{1/3} \biggl[ 1 +{1
   \over 5} \biggl( {\O^2 \over 2M_{\rm PPN}} \biggr)^{2/3} (2a_{1 {\rm
     PPN}}^2 -a_{2 {\rm PPN}}^2 -a_{3 {\rm PPN}}^2) \nonumber \\
  & &+{1 \over c^2} \biggl\{ -{11 M_{\rm PPN} \over 6} \biggl( {\O^2
    \over 2M_{\rm PPN}} \biggr)^{1/3}
  -{22 \pi \r A_{0 {\rm PPN}} \over 75} \biggl(
   {\O^2 \over 2M_{\rm PPN}} \biggr)^{2/3} (2a_{1 {\rm PPN}}^2 -a_{2
     {\rm PPN}}^2 -a_{3 {\rm PPN}}^2) \nonumber \\
  & &-{M_{\rm PPN} \over 30} \biggl(
   {\O^2 \over 2M_{\rm PPN}} \biggr) (2 a_{1 {\rm PPN}}^2 +5 a_{2 {\rm
       PPN}}^2 +6 a_{3 {\rm PPN}}^2) \biggr\} \biggr]. \label{sepppn}
\eeqa
Also if we use $M_*$, the orbital separation is
written as
\beqa
  R &=& \biggl( {2 M_{\ast} \over \O^2} \biggr)^{1/3} \biggl[ 1 +{1
   \over 5} \biggl( {\O^2 \over 2M_{\ast}} \biggr)^{2/3} (2a_{1 \ast}^2
   -a_{2 \ast}^2 -a_{3 \ast}^2) +{1 \over c^2} \biggl\{ -{2 \pi \r A_{0 
   \ast} \over 15} -{11 M_{\ast} \over 6} \biggl( {\O^2 \over
   2M_{\ast}} \biggr)^{1/3} \nonumber \\
  & &-{8 \pi \r A_{0 \ast} \over 25} \biggl(
   {\O^2 \over 2M_{\ast}} \biggr)^{2/3} (2a_{1 \ast}^2 -a_{2 \ast}^2
    -a_{3 \ast}^2) -{M_{\ast} \over 10} \biggl( {\O^2 \over
   2M_{\ast}} \biggr) (a_{2 \ast}^2 +2 a_{3 \ast}^2) \biggr\}
 \biggr]. \label{sepcons}
\eeqa
Substituting these expressions into the equation of the energy loss 
rate, we have
\beqa
  {dE \over dt} &=&-{2 \over 5c^5} (2 M_{\rm PPN} \O)^{10/3} \Biggl[ 1+
  {(2 M_{\rm PPN} \O)^{4/3} \over 5 M_{\rm PPN}^2} \biggl\{ 2(a_{1 {\rm
      PPN}}^2 -a_{2 {\rm PPN}}^2) +2a_{1 {\rm  PPN}}^2 -a_{2 {\rm
      PPN}}^2 -a_{3 {\rm PPN}}^2 \biggr\} \nonumber \\
  & &+{1 \over c^2} \biggl[ -{373 \over 84} (2 M_{\rm PPN} \O)^{2/3}
  \nonumber \\
  & &\hspace{30pt} -{2 \pi \r A_{0 {\rm PPN}} \over 525 M_{\rm PPN}^2}
  (2 M_{\rm PPN} \O)^{4/3} \biggl\{ 160(a_{1 {\rm PPN}}^2 -a_{2 {\rm
      PPN}}^2) +77(2 a_{1 {\rm PPN}}^2 -a_{2 {\rm PPN}}^2 -a_{3 {\rm
      PPN}}^2) \biggr\} \nonumber \\
  & &\hspace{30pt} -{(2 M_{\rm PPN} \O)^2 \over 840 M_{\rm PPN}^2} (2602
  a_{1 {\rm PPN}}^2 -1697 a_{2 {\rm PPN}}^2 -443 a_{3 {\rm PPN}}^2) \biggr]
  \Biggr] , \label{dEdtomegappn}
\eeqa
or
\beqa
  {dE \over dt} &=& -{2 \over 5c^5} (2 M_{\ast} \O)^{10/3} \Biggl[ 1+
  {(2 M_{\ast} \O)^{4/3} \over 5 M_{\ast}^2} \biggl\{ 2(a_{1 \ast}^2 -a_{2
    \ast}^2) +2a_{1 \ast}^2 -a_{2 \ast}^2 -a_{3 \ast}^2 \biggr\}
  \nonumber \\
  & &+{1 \over c^2} \biggl[ -{4
    \pi \r A_{0 \ast} \over 3} -{373 \over 84} (2 M_{\ast} \O)^{2/3}
  -{2 \pi \r A_{0 \ast} \over 175 M_{\ast}^2} (2 M_{\ast} \O)^{4/3}
  \biggl\{ 100(a_{1 \ast}^2 -a_{2 \ast}^2) +49(2a_{1 \ast}^2 -a_{2
    \ast}^2 -a_{3 \ast}^2) \biggr\} \nonumber \\
  & &\hspace{30pt}-{(2 M_{\ast}
    \O)^2 \over 840 M_{\ast}^2} (2532 a_{1 \ast}^2 -1767 a_{2 \ast}^2
  -443 a_{3 \ast}^2) \biggr] \Biggr]. \label{dEdtomegacons}
\eeqa
On the other hand, in the point particle case, the energy loss rate is
written as
\beqa
  {dE \over dt} = -{2 \over 5c^5} (2 M_{\rm PPN} \O)^{10/3} \biggl[ 1+
  {1 \over c^2} \biggl\{ -{373 \over 84} (2 M_{\rm PPN} \O)^{2/3} -{(2
    M_{\rm PPN} \O)^2 \over 5 M_{\rm PPN}^2} (a_{1 {\rm PPN}}^2 +a_{2
    {\rm PPN}}^2) \biggr\} \biggr]. \label{dEdtomegapoint}
\eeqa
Comparing the last term of Eq. (\ref{dEdtomegappn}) with that of Eq.
(\ref{dEdtomegapoint}), we find that they do not agree. A part of 
the reason seems due to the fact that the 1PN quadrupole dependent term 
really appears in the last terms of Eq. (\ref{dEdtomegappn}). 
However, we cannot still explain the 
disagreement completely because there remains terms except for 
the SO coupling one even if we 
take a spherical limit $a_{1 {\rm PPN}}=a_{2 {\rm PPN}}=a_{3 {\rm PPN}}$. 
Then, what is the reason why the terms remain? 
We guess that this is due to the 
imperfection of the definition of gravitational mass. 
We adopt the PPN mass as gravitational mass of each star to write 
the equation of the energy loss rate, but the PPN mass can 
renormalize only the self-gravity and spin into mass. In the context of 
this paper, we need a conserved gravitational mass which 
renormalizes the quadrupole term produced by the tidal force of the
companion star.
As far as we know, however, no one has proposed such a mass. 
Nevertheless, $M_{\rm PPN}$ is not expected to be different from ``true 
gravitational mass'' $M_{\rm true}$ so much because 
\beq
M_{\rm true}=M_{\rm PPN}\biggl[1+O\biggl( {Ma_i^2 \over c^2 R^3}\biggr)
\biggr]. 
\eeq
Hence, we regard $M_{\rm PPN}$ as an 
approximate gravitational mass in the following. 

\subsection{The angular momentum loss rate}

Next, we calculate the angular momentum loss rate. As shown by
Ostriker and Gunn\cite{OG}, the angular momentum loss rate has the
relation
\beqa
  {dJ^3 \over dt} ={1 \over \O} {dE \over dt}
\eeqa
for any stable, uniformly rotating, equilibrium configuration. Then, the
angular momentum loss rate is written as
\beqa
  {dJ^3 \over dt} &=&-{4 M_{\rm PPN} \over 5c^5} (2 M_{\rm PPN}
  \O)^{7/3} \Biggl[ 1+
  {(2 M_{\rm PPN} \O)^{4/3} \over 5 M_{\rm PPN}^2} \biggl\{ 2(a_{1 {\rm
      PPN}}^2 -a_{2 {\rm PPN}}^2) +2a_{1 {\rm  PPN}}^2 -a_{2 {\rm
      PPN}}^2 -a_{3 {\rm PPN}}^2 \biggr\} \nonumber \\
  & &+{1 \over c^2} \biggl[ -{373 \over 84} (2 M_{\rm PPN} \O)^{2/3}
  \nonumber \\
  & &-{2 \pi \r A_{0 {\rm PPN}} \over 525 M_{\rm PPN}^2} (2 M_{\rm PPN}
  \O)^{4/3} \biggl\{ 160(a_{1 {\rm PPN}}^2 -a_{2 {\rm PPN}}^2) +77(2
  a_{1 {\rm PPN}}^2 -a_{2 {\rm PPN}}^2 -a_{3 {\rm PPN}}^2) \biggr\}
  \nonumber \\
  & &-{(2 M_{\rm PPN} \O)^2 \over 840 M_{\rm PPN}^2} (2602 a_{1 {\rm
      PPN}}^2 -1697 a_{2 {\rm PPN}}^2 -443 a_{3 {\rm PPN}}^2) \biggr]
  \Biggr], \label{djdtppn}
\eeqa
or
\beqa
  {dJ^3 \over dt} &=& -{4 M_{\ast} \over 5c^5} (2 M_{\ast} \O)^{7/3}
  \Biggl[ 1+
  {(2 M_{\ast} \O)^{4/3} \over 5 M_{\ast}^2} \biggl\{ 2(a_{1 \ast}^2 -a_{2
    \ast}^2) +2a_{1 \ast}^2 -a_{2 \ast}^2 -a_{3 \ast}^2 \biggr\}
  \nonumber \\
  & &+{1 \over c^2} \biggl[ -{4
    \pi \r A_{0 \ast} \over 3} -{373 \over 84} (2 M_{\ast} \O)^{2/3}
  -{2 \pi \r A_{0 \ast} \over 175 M_{\ast}^2} (2 M_{\ast} \O)^{4/3}
  \biggl\{ 100(a_{1 \ast}^2 -a_{2 \ast}^2) +49(2a_{1 \ast}^2 -a_{2
    \ast}^2 -a_{3 \ast}^2) \biggr\} \nonumber \\
  & &\hspace{30pt}-{(2 M_{\ast}
    \O)^2 \over 840 M_{\ast}^2} (2532 a_{1 \ast}^2 -1767 a_{2 \ast}^2
  -443 a_{3 \ast}^2) \biggr] \Biggr]. \label{djdtcons}
\eeqa
We will also show the direct calculation in Appendix A.

\section{Numerical results}

In this section, we calculate the energy loss rate along the equilibrium 
sequence of the binary. The method is as follows.

\noindent
(i) Using the equilibrium equations for a corotating binary, we
construct the sequence of the binary. The equations are written as
\beqa
  -{P_0 \over \r} &=& -\pi \r a_1^2 A_1 +{M \over R^3} a_1^2 +{\O_{\rm
      N}^2 \over 2} a_1^2, \\
  -{P_0 \over \r} &=& -\pi \r a_2^2 A_2 -{M \over 2R^3} a_2^2 +{\O_{\rm
      N}^2 \over 2} a_2^2, \\
  -{P_0 \over \r} &=& -\pi \r a_3^2 A_3 -{M \over 2R^3} a_3^2,
\eeqa
where $\O_{\rm N}$ denotes the Newtonian angular velocity written by
\beqa
  \O_{\rm N}^2 ={2 M \over R^3} +{18 \bI_{11} \over R^5}.
\eeqa
At this stage, we determine
$\t{R}=R/a_1,~\a_2=a_2/a_1$ and $\a_3=a_3/a_1$.

\noindent
(ii) Inserting the equilibrium figures into the equation of the orbital
angular velocity of 1PN order
\beqa
  \O &=& \biggl( {2 M_{\ast} \over R^3} \biggr)^{1/2} \biggl[ 1+ {3
    \over 10 R^2} (2a_{1 \ast}^2 -a_{2 \ast}^2 -a_{3 \ast}^2) -{1 \over
    c^2} \biggl\{ {\pi \r A_{0 \ast} \over 5} +{11 M_{\ast} \over 4R}
  \nonumber \\
  & &\hspace{50pt} +{29 \pi \r A_{0 \ast} \over 50 R^2} (2a_{1 \ast}^2
  -a_{2 \ast}^2 -a_{3 \ast}^2) +{M_{\ast} \over 40R^3} (154a_{1 \ast}^2
  -71a_{2 \ast}^2 -65a_{3 \ast}^2) \biggr\} \biggr]
\eeqa
and the equation of the total energy of the binary\cite{TS},
we determine the initial and final points. Here, we
regard the point which satisfies the condition $\O/\pi=10$Hz as the
initial one and the point which satisfies $\O/\pi=1000$Hz or reaches the
energy minimum (ISCO)\footnote{In the corotating binary case, we call it 
the innermost stable corotating circular orbit (ISCCO)\cite{TS}.} as
the final one. 

\noindent
(iii) Substituting the equilibrium figures into
Eq. (\ref{dEdtomegappn}), we are able to have the energy loss rate.
However, Eq. (\ref{dEdtomegappn}) is not written by conserved
quantities. Then, we express Eq. (\ref{dEdtomegappn}) as
\beqa
  \bigl| \bar{\dot{E}} \bigr| =
  1 + \bar{\dot{E}}_{\rm N-finite} +\bar{\dot{E}}_{\rm 1PN} +
  \bar{\dot{E}}_{\rm 1PN-finite}
\eeqa
where
\beqa
  \bar{\dot{E}} &\equiv& {dE \over dt} \bigg/ \biggl( {dE \over dt}
  \biggr)_{\rm N}, \\
  \biggl( {dE \over dt} \biggr)_{\rm N} &=& -{2 \over 5c^5} ( 2 M_{\rm
    PPN} \O)^{10/3}, \\
  \bar{\dot{E}}_{\rm N-finite} &=& {1 \over 5C_{\rm s}^2} \biggl(
  {2M_{\ast} \over c^2} {\O \over c} \biggr)^{4/3} {1 \over (\a_2
    \a_3)^{2/3}} \Bigl\{ 2(1-\a_2^2) +2-\a_2^2 -\a_3^2 \Bigr\},
  \label{dotEnf} \\
  \bar{\dot{E}}_{\rm 1PN} &=& -{373 \over 84} \biggl( {2M_{\ast} \over
    c^2} {\O \over c} \biggr)^{2/3}, \label{dotE1pn} \\
  \bar{\dot{E}}_{\rm 1PN-finite} &=& -{1 \over 350C_{\rm s}} \biggl(
  {2M_{\ast} \over c^2} {\O \over c} \biggr)^{4/3} {\tilde{A}_0 \over
    (\a_2 \a_3)^{4/3}} \Bigl\{ 160(1-\a_2^2) +77(2-\a_2^2 -\a_3^2)
    \Bigr\}. \label{dotE1pnf}
\eeqa
In these equations, we use the compactness parameter
\beqa
  C_{\rm s} \equiv {M_{\ast} \over c^2 a_{\ast}}.
\eeqa
In calculation of the energy loss rate, we neglect the last term of
Eq. (\ref{dEdtomegappn}), because this term includes the ambiguity of the
definition of mass and we cannot separate the SO coupling term from it
as mentioned in the last of subsection \ref{subsecelr}.

We repeat this procedure changing the compactness of the star
$M_{\ast}/c^2 a_{\ast}$ and the conserved proper mass $M_{\ast}$.

Figures 1(a)--3(c) show the results we have using this procedure. It is
found from these figures that the finite size effect of the 1PN order is
only a few factor less than that of the Newtonian order. We can explain
this feature by comparing Eq. (\ref{dotEnf}) with
Eq. (\ref{dotE1pnf}). Essentially, the difference between these
equations is $O(C_{\rm s})$. Therefore,
the finite size effect of the 1PN order becomes $O(C_{\rm s})$
less than that of the Newtonian order.

There is another feature to mention specially. Comparing
Eq. (\ref{dotEnf}) with Eq. (\ref{dotE1pnf}), it is found that the
contribution from the 1PN order has the opposite sign of that from the
Newtonian order. We explain this fact as follows: Including the 1PN
order terms, the self-gravity of each star of binary becomes stronger
than that in the Newtonian gravity. Then, the stars become more
compact, and it is more difficult to deform them. This leads to the
decrease of the finite size effect. Moreover, it is found from these
figures that when we increase the compactness parameter fixing $M_*$, 
both the finite size effects of the Newtonian order and
the 1PN one decrease, and also the difference between their absolute
values decreases. The explanation for these behaviors is the same as the 
above one, i.e., the self-gravity of each star becomes stronger.

Finally, we discuss on the inclinations of the lines in the
figures. First, the inclination of the 1PN line becomes 2/3 in the
log-log plot figures because of Eq. (\ref{dotE1pn}). Next, we can see from 
the figures that the inclination of the finite size effect lines is 
$\sim 3.4$. This is because the deviation from the spherically symmetric 
star is made by the tidal force $\sim M/R^3$. We find from
Eq. (\ref{1PNomega}) that $M/R^3$ is equal to $\O^2$. Then, combining
this fact with Eq. (\ref{dotEnf}) or (\ref{dotE1pnf}), we can conclude
the inclination of the finite size effect lines becomes $\sim 10/3$.

In Table I, we show the results of the initial
and final orbits. Note that the frequencies of the final orbits are
1000Hz in the cases of $(M_{\ast}/M_{\odot},~M_{\ast}/c^2
a_{\ast})=(1.4,~0.20)$, $(1.4,~0.25)$, $(1.6,~0.25)$, and $(1.8,~0.25)$, 
while in the other cases, the ISCCOs are the final orbits.
Also we present the PPN mass (gravitational mass) at infinity which is
written as
\beqa
  M_{\rm PPN,inf} =M_{\ast} \biggl( 1-{3 \over 5} C_{\rm s} \biggr)
  ~~~~~(R \ra \infty).
\eeqa

\section{summary and discussion}

In this paper, we have analytically calculated the energy and
the angular momentum loss rates. The conclusions are as follows.

\noindent
(i) The leading finite size term of the 1PN order in $dE/dt$ reduces 
$(dE/dt)_{\rm N-finite}$ by $\sim 40\%$ which is $2C_{\rm s}$ times as
large as that of the Newtonian order. 

\noindent
(ii) The reason of the weak convergence of the finite size terms is that
both of them are in the same order when we see Eq. (\ref{dEdtomegappn}) in
the viewpoint of the series of $\O$. However, the 1PN term is of 
$O(C_{\rm s})$ higher than the Newtonian one in the viewpoint of the PN
approximation. Therefore, the 1PN term is only by a factor of 
$O(C_{\rm s})$ smaller than the Newtonian one.

\noindent
(iii) The leading finite size term of the 1PN order has the opposite
sign of that of the Newtonian order. This is because the 1PN order terms 
make the self-gravity of the star stronger, then it is difficult to
deform the star. Therefore, the finite size effects on the energy loss
rate decrease.

Next, we discuss the definition of gravitational mass of 
each star. In this paper, we use the 
PPN mass as the gravitational mass 
in order to compare the equations of the energy loss rate
with those of the point particle case. However, the PPN mass is
appropriate only for the point particle case (i.e.,  
as long as we ignore the tidal forces on each star of binary)\cite{Will}. 
In the context of present paper, 
we should use a gravitational mass which conserves and 
appropriately renormalizes 
the effect of the finite size of stars even if the tidal
force exists. Unfortunately, no one has proposed such a mass 
as far as we know. To define such gravitational 
mass will be an unresolved problem. 

Finally, we point out a problem with regard to 
the back reaction by gravitational radiation as follows. 
In this paper, we have calculated the total gravitational wave luminosity. 
However, it is not trivial what fraction of it contributes to the 
back reaction to $\Omega$ because the structure of each star
(i.e., $a_i$ in the incompressible case) is also affected due to 
the radiation reaction. This situation is in contrast to that for 
the point particle case where we only need to consider the radiation 
reaction to $\Omega$. 
To know the change rate of $\Omega$ and $a_i$ in the 1PN order, 
we need to solve the hydrodynamic equation in the 1PN 
order including the 3.5PN radiation reaction terms. Here, 
we note that from the observational point of view, 
we need not the total luminosity, but the back reaction of $\Omega$. 
Hence, to resolve this problem seems one of the important issues 
in order to study the late time evolution of BNS's.

\acknowledgments 

We thank H. Asada, L. Blanchet and K. Nakao for useful discussions and
also K.T. would like to thank T. Nakamura and H. Sato for helpful
comments and continuous encouragement. This work was in part supported
by a Grant-in-Aid of Ministry of Education, Culture, Science and Sports
(Nos. 08237210, 09740336 and 08NP0801).

\vskip 5mm

\appendix
\section{The direct calculation of the angular momentum loss rate}

In the case of the identical star binary, $M_{ijk}$ and 
$S_{ij}$ are zero. Therefore, the angular
momentum loss is written as
\beqa
  {dJ^i \over dt} =- {2 \over 5 c^5} \epsilon_{ijk} \biggl[
  (M_{jl}^{(2)})_{\rm tot} (M_{kl}^{(3)})_{\rm tot} \biggr].
\eeqa

For the quadrupole moment, the contributions from star 2 are the same as
those from star 1. Then, the total quadrupole moments double those of
star 1.

In the case we take in this paper, the binary rotates around the $X_3$
axis. Then, the angular momentum loss has only $i=3$ component and is
written as
\beqa
  {dJ^3 \over dt} &=&- {2 \over 5 c^5} \epsilon_{3jk} \biggl[
  (2D_{jl}^{(2)}) (2D_{kl}^{(3)}) \nonumber \\
  & &\hspace{45pt}+{1 \over c^2} \biggl\{
  (2D_{jl}^{(2)}) \biggl( 2V_{kl}^{(3)} +2U_{kl}^{ 1 \ra 1 (3)}
  +2U_{kl}^{ 2 \ra 1 (3)} +2P_{kl}^{(3)} +2Q_{kl}^{(3)} +2R_{kl}^{(3)}
  \biggr) \nonumber \\
  & &\hspace{50pt}+(2D_{kl}^{(3)}) \biggl( 2V_{jl}^{(2)} +2U_{jl}^{ 1
    \ra 1 (2)} +2U_{jl}^{ 2 \ra 1 (2)} +2P_{jl}^{(2)} +2Q_{jl}^{(2)}
  +2R_{jl}^{(2)} \biggr) \biggr\} \biggr].
\eeqa
When we use the explicit forms for the quadrupole moments, the angular
momentum loss is rewritten as
\beqa
  {dJ^3 \over dt} &=& - {8 \over 5 c^5} \biggl[ 2D_{11}^{(2)}
  D_{12}^{(3)} -2D_{12}^{(2)} D_{11}^{(3)} \nonumber \\
  & &\hspace{30pt}+{1 \over c^2} \biggl\{
  -D_{12}^{(2)} \biggl( (V_{11}-V_{22})^{(3)} +(U_{11}^{1 \ra 1}-
    U_{22}^{1 \ra 1})^{(3)} + (U_{11}^{2 \ra 1}-U_{22}^{2 \ra 1})^{(3)}
  \nonumber \\
  & &\hspace{100pt}+(P_{11}-P_{22})^{(3)} +(Q_{11}-Q_{22})^{(3)}
  +(R_{11}-R_{22})^{(3)} \biggr) \nonumber \\
  & &\hspace{60pt} +2D_{11}^{(2)} \biggl( V_{12}^{(3)}
  +U_{12}^{1 \ra 1 (3)} +U_{12}^{2 \ra 1 (3)} +P_{12}^{(3)} +Q_{12}^{(3)}
    +R_{12}^{(3)} \biggr) \nonumber \\
  & &\hspace{60pt}+D_{12}^{(3)} \biggl( (V_{11}-V_{22})^{(2)}
    +(U_{11}^{1 \ra 1}- U_{22}^{1 \ra 1})^{(2)} + (U_{11}^{2 \ra
      1}-U_{22}^{2 \ra 1})^{(2)} \nonumber \\
  & &\hspace{100pt} +(P_{11}-P_{22})^{(2)} +(Q_{11}-Q_{22})^{(2)}
  +(R_{11}-R_{22})^{(2)} \biggr) \nonumber \\
  & &\hspace{60pt} -2D_{11}^{(3)} \biggl( V_{12}^{(2)}
  +U_{12}^{1 \ra 1 (2)} +U_{12}^{2 \ra 1 (2)} +P_{12}^{(2)} +Q_{12}^{(2)}
    +R_{12}^{(2)} \biggr) \biggr\} \biggr], \nonumber \\
\eeqa 
where we use the relation $D_{22}^{(i)} =-D_{11}^{(i)}$.

The contribution from the Newtonian order is
\beqa
  2D_{11}^{(2)} D_{12}^{(3)} -2D_{12}^{(2)}
  D_{11}^{(3)} = 16 \O^5 \hat{D}^2.
\eeqa

The contributions from the 1PN order are written as follows:
\beqa
  -D_{12}^{(2)} (V_{11}-V_{22})^{(3)} +2D_{11}^{(2)} V_{12}^{(3)}
  &=& D_{12}^{(3)} (V_{11}-V_{22})^{(2)} -2D_{11}^{(3)} V_{12}^{(2)}
  \nonumber \\
  &=& 32 \O^7 \hat{D} \hat{V}, \\
  -D_{12}^{(2)}(U_{11}^{1 \ra 1}-U_{22}^{1 \ra 1})^{(3)} +
  2D_{11}^{(2)} U_{12}^{1 \ra 1 (3)}
  &=& D_{12}^{(3)} (U_{11}^{1 \ra 1}- U_{22}^{1 \ra 1})^{(2)}
  -2D_{11}^{(3)} U_{12}^{1 \ra 1 (2)} \nonumber \\
  &=& 32 \O^5 \hat{D} \hat{U}^{1 \ra 1}, \\
  -D_{12}^{(2)} (U_{11}^{2 \ra 1}-U_{22}^{2 \ra 1})^{(3)} +
  2D_{11}^{(2)} U_{12}^{2 \ra 1 (3)}
  &=& D_{12}^{(3)} (U_{11}^{2 \ra 1}-U_{22}^{2 \ra 1})^{(2)} -
  2D_{11}^{(3)} U_{12}^{2 \ra 1 (2)} \nonumber \\
  &=& 32 \O^5 \hat{D} \hat{U}^{2 \ra 1}, \\
  -D_{12}^{(2)} (P_{11}-P_{22})^{(3)} + 2D_{11}^{(2)} P_{12}^{(3)}
  &=& D_{12}^{(3)} (P_{11}-P_{22})^{(2)} -2D_{11}^{(3)} P_{12}^{(2)}
  \nonumber \\
  &=& 48 \O^5 \hat{D} \hat{P}, \\
  -D_{12}^{(2)} (Q_{11}-Q_{22})^{(3)} + 2D_{11}^{(2)} Q_{12}^{(3)}
  &=& D_{12}^{(3)} (Q_{11}-Q_{22})^{(2)} -2D_{11}^{(3)} Q_{12}^{(2)}
  \nonumber \\
  &=& -{32 \over 7} \O^7 \hat{D} \hat{Q}, \\
  -D_{12}^{(2)} (R_{11}-R_{22})^{(3)} + 2D_{11}^{(2)} R_{12}^{(3)}
  &=& D_{12}^{(3)} (R_{11}-R_{22})^{(2)} -2D_{11}^{(3)} R_{12}^{(2)}
  \nonumber \\
  &=& -{256 \over 21} \O^7 \hat{D} \hat{Q}.
\eeqa

Gathering these terms, we have the angular momentum loss rate as
\beqa
  {dJ^3 \over dt} &=& -{8 \over 5 c^5} \O_{{\rm N}}^5 R^4 M^2 \Biggl[
  1+ {8 \over 5R^2} (a_1^2 -a_2^2) +O(R^{-4}) \nonumber \\
  & & +{1 \over c^2} \biggl[ 9 \pi \r A_0 -{113 M \over 168 R} +{ \pi \r 
    A_0 \over R^2} \biggl\{ {296 \over 21} (a_1^2-a_2^2) -{1 \over 5}
  (2a_1^2 -a_2^2 -a_3^2) \biggl\} \nonumber \\
  & &\hspace{30pt}+{M \over 840 R^3} (358 a_1^2 +161
  a_2^2 -2465 a_3^2) +O(R^{-4}) \biggr] \Biggr], \\
  &=& -{4 M \over 5 c^5} \biggl( {2M \over R} \biggr)^{7/2} \Biggl[
  1+ {1 \over R^2} \biggl\{ {8 \over 5} (a_1^2 -a_2^2) + {3 \over 2}
  (2a_1^2 -a_2^2 -a_3^2) \biggr\} +O(R^{-4}) \nonumber \\
  & &+{1 \over c^2} \biggl[ 9 \pi \r A_0 -{113 M \over 168 R} +{ \pi \r
   A_0 \over R^2} \biggl\{ {296 \over 21} (a_1^2-a_2^2) +{133 \over 10}
  (2a_1^2 -a_2^2 -a_3^2) \biggl\} \nonumber \\
  & &\hspace{30pt} -{M \over 1680 R^3} (2674 a_1^2 - 2017
  a_2^2 +3235 a_3^2) +O(R^{-4}) \biggr] \Biggr]. \label{angmomloss}
\eeqa

As in the case of the energy loss rate, there remains the internal
structure dependent term in the limits $a_i/R \rightarrow 0$ when we
rewrite the angular momentum loss rate using $M_*$ as
\beqa
  {dJ^3 \over dt} &=& -{4 M_{\ast} \over 5 c^5} \biggl( {2M_{\ast} \over
    R} \biggr)^{7/2} \Biggl[
  1+ {1 \over R^2} \biggl\{ {8 \over 5} (a_1^2 -a_2^2) + {3 \over 2}
  (2a_1^2 -a_2^2 -a_3^2) \biggr\} +O(R^{-4}) \nonumber \\
  & &+{1 \over c^2} \biggl[ -{9 \pi \rho A_0 \over 5} -{1285 M_{\ast}
    \over 84 R} -{ \pi \r
  A_0 \over 5 R^2} \biggl\{ {1672 \over 105} (a_1^2-a_2^2) + {29 \over
  2} (2a_1^2 -a_2^2 -a_3^2) \biggl\} \nonumber \\
  & &\hspace{30pt} -{M_{\ast} \over 840 R^3} (64274 a_1^2 -41171
  a_2^2 -19645 a_3^2) +O(R^{-4}) \biggr] \Biggr], \label{dJdtcons} \\
  &=& -{4 M_{\ast} \over 5 c^5} \biggl( {2M_{\ast} \over R_{\ast}}
  \biggr)^{7/2} \Biggl[ 1+ {1 \over R_{\ast}^2} \biggl\{ {8 \over 5}
  (a_{1 \ast}^2 -a_{2 \ast}^2) + {3 \over 2} (2a_{1 \ast}^2 -a_{2
    \ast}^2 -a_{3 \ast}^2) \biggr\} +O(R_{\ast}^{-4}) \nonumber \\
  & &+{1 \over c^2} \biggl[ -{9 \pi \rho A_{0 \ast} \over 5} -{1285
    M_{\ast} \over 84 R_{\ast}} -{ \pi \r A_{0 \ast} \over 5 R_{\ast}^2}
  \biggl\{ {3016 \over 105} (a_{1 \ast}^2 -a_{2 \ast}^2) + {53 \over
  2} (2a_{1 \ast}^2 -a_{2 \ast}^2 -a_{3 \ast}^2) \biggl\} \nonumber \\
  & &\hspace{30pt} -{M_{\ast} \over 840 R_{\ast}^3} (74998 a_{1 \ast}^2
  -46813 a_{2 \ast}^2 -22375 a_{3 \ast}^2) +O(R_{\ast}^{-4}) \biggr]
  \Biggr].
\eeqa
However, when we substitute the PPN mass into Eq. (\ref{angmomloss}), 
we have
\beqa
  {dJ^3 \over dt} &=& -{4 M_{\rm PPN} \over 5 c^5} \biggl( {2M_{\rm
        PPN} \over R} \biggr)^{7/2} \Biggl[
  1+ {1 \over R^2} \biggl\{ {8 \over 5} (a_1^2 -a_2^2) + {3 \over 2}
  (2a_1^2 -a_2^2 -a_3^2) \biggr\} +O(R^{-4}) \nonumber \\
  & &+{1 \over c^2} \biggl[ -{1285 M_{\rm PPN} \over 84 R} -{ \pi \r
  A_0 \over 5R^2} \biggl\{ {32 \over 21} (a_1^2-a_2^2) + (2a_1^2 -a_2^2
  -a_3^2) \biggl\} \nonumber \\
  & &\hspace{30pt} -{M_{\rm PPN} \over 168 R^3} (13006 a_1^2 -8083
  a_2^2 -3929 a_3^2) +O(R^{-4}) \biggr] \Biggr], \label{dJdtppn} \\
  &=& -{4 M_{\rm PPN} \over 5 c^5} \biggl( {2M_{\rm PPN} \over R_{\rm
      PPN}} \biggr)^{7/2} \Biggl[ 1+ {1 \over R_{\rm PPN}^2} \biggl\{ {8
    \over 5} (a_{1 {\rm PPN}}^2 -a_{2 {\rm PPN}}^2) + {3 \over 2} (2a_{1
    {\rm PPN}}^2 -a_{2 {\rm PPN}}^2 -a_{3 {\rm PPN}}^2) \biggr\}
  \nonumber \\
  & &\hspace{130pt} +O(R_{\rm PPN}^{-4}) \nonumber \\
  & &+{1 \over c^2} \biggl[ -{1285 M_{\rm PPN} \over 84 R_{\rm PPN}} -
  { \pi \r A_{0 {\rm PPN}} \over 5R_{\rm PPN}^2} \biggl\{ {256 \over 21}
  (a_{1 {\rm PPN}}^2-a_{2 {\rm PPN}}^2) + 11(2a_{1 {\rm PPN}}^2 -a_{2
    {\rm PPN}}^2 -a_{3 {\rm PPN}}^2) \biggl\} \nonumber \\
  & &\hspace{30pt} -{M_{\rm PPN} \over 840 R_{\rm PPN}^3} (75754 a_{1
    {\rm PPN}}^2 -46057 a_{2 {\rm PPN}}^2 -22375 a_{3 {\rm PPN}}^2)
  +O(R_{\rm PPN}^{-4}) \biggr] \Biggr],
\eeqa
and this expression does not depend on the internal structure of the star.

If we substitute Eqs. (\ref{sepppn}) and/or (\ref{sepcons}) into
Eqs. (\ref{dJdtppn}) and/or (\ref{dJdtcons}), we have the same equations
as (\ref{djdtppn}) and/or (\ref{djdtcons}).

\vspace{1.cm}

\begin{center}
  {\large TABLE CAPTIONS}
\end{center}

\vskip 5mm

\noindent
Table I. The coordinate orbital separation $R_{\ast}/a_{\ast}$, axial
ratios $a_2/a_1$ and $a_3/a_1$, and the frequency of gravitational wave
$\O/\pi [{\rm Hz}]$ of the initial and final orbits.

\vskip 1cm

\begin{table}
 \begin{center}
  \begin{tabular}{ccc|cccc|cccc}
   \multicolumn{3}{c|}{}&
   \multicolumn{4}{c|}{Initial}& \multicolumn{4}{c}{Final} \\ \hline
    $M_{\ast}/M_{\odot}$&$M_{\ast}/c^2 a_{\ast}$&
    $M_{\rm PPN,inf}/M_{\odot}$&$R_{\ast}/a_{\ast}$&
    $a_2/a_1$&$a_3/a_1$&$\O/\pi[{\rm Hz}]$&$R_{\ast}/a_{\ast}$&$a_2/a_1$&
    $a_3/a_1$&$\O/\pi[{\rm Hz}]$ \\ \hline
      &0.15&1.274&48.9478&0.999983&0.999971&10.00
             &2.6511&0.888010&0.837826&706.08 \\
   1.4&0.20&1.232&62.9341&0.999993&0.999989&10.00
             &2.6278&0.918423&0.877899&1000.0 \\
      &0.25&1.190&75.2153&0.999997&0.999995&10.00
             &3.1321&0.965573&0.945395&1000.0 \\ \hline
      &0.15&1.456&44.7585&0.999977&0.999962&10.00
             &2.6511&0.888010&0.837826&617.82 \\
   1.6&0.20&1.408&57.5478&0.999991&0.999986&10.00
             &2.4554&0.898344&0.851178&959.53 \\
      &0.25&1.360&68.7777&0.999996&0.999993&10.00
             &2.8328&0.953794&0.927832&1000.0 \\ \hline
      &0.15&1.638&41.3604&0.999971&0.999952&10.00
             &2.6511&0.888010&0.837826&549.17 \\
   1.8&0.20&1.584&53.1789&0.999989&0.999982&10.00
             &2.4554&0.898344&0.851178&852.92 \\
      &0.25&1.530&63.5560&0.999995&0.999992&10.00
            &2.5904&0.939745&0.907524&1000.0
  \end{tabular}
 \end{center}
 \caption{}
 \label{table2}
\end{table}%

\vskip 1cm

\begin{center}
  {\large FIGURE CAPTIONS}
\end{center}

\vskip 5mm

\noindent
Figs.1. The energy loss rates relative to that of the Newtonian
order are shown as a function of the frequency of gravitational waves
$f_{\rm GW} =\O/\pi$. The sequences are calculated from the initial
orbit (10Hz) to the
final one (1000Hz or ISCCO). In these figures, we fix the conserved 
proper mass 
as $M_{\ast}=1.4M_{\odot}$ and change the compactness parameter as
$C_{\rm s}=0.15$ (a), 0.2 (b), 0.25 (c). Solid, dotted and dashed lines
denote contributions from the finite size effect of the Newtonian order,
that of the 1PN order and the 1PN order terms, respectively. The
long-dashed line denotes the contribution from the Newtonian order term 
which is unity in these figures. In fig.1(a), we present the location of 
the energy minimum point (ISCCO) using the solid vertical line.

\vskip 5mm

\noindent
Figs.2. The conventions are the same as those in figs.1, but for $M_*$. 
Here, we fix it as $M_{\ast}=1.6M_{\odot}$ and change
the compactness parameter as $C_{\rm s}=0.15$ (a), 0.2 (b), 0.25 (c).
In fig.2(a) and 2(b), we present the location of the energy minimum
point (ISCCO) using the solid vertical line.

\vskip 5mm

\noindent
Figs.3. The conventions are the same as those in figs.1, but for $M_*$. 
Here, we fix it as $M_{\ast}=1.8M_{\odot}$ and change
the compactness parameter as $C_{\rm s}=0.15$ (a), 0.2 (b), 0.25 (c).
In fig.3(a) and 3(b), we present the location of the energy minimum
point (ISCCO) using the solid vertical line.

\newpage

\begin{figure}[ht]
  \vspace{1cm}
  \centerline{\epsfysize 15cm \epsfxsize 15cm \epsfbox{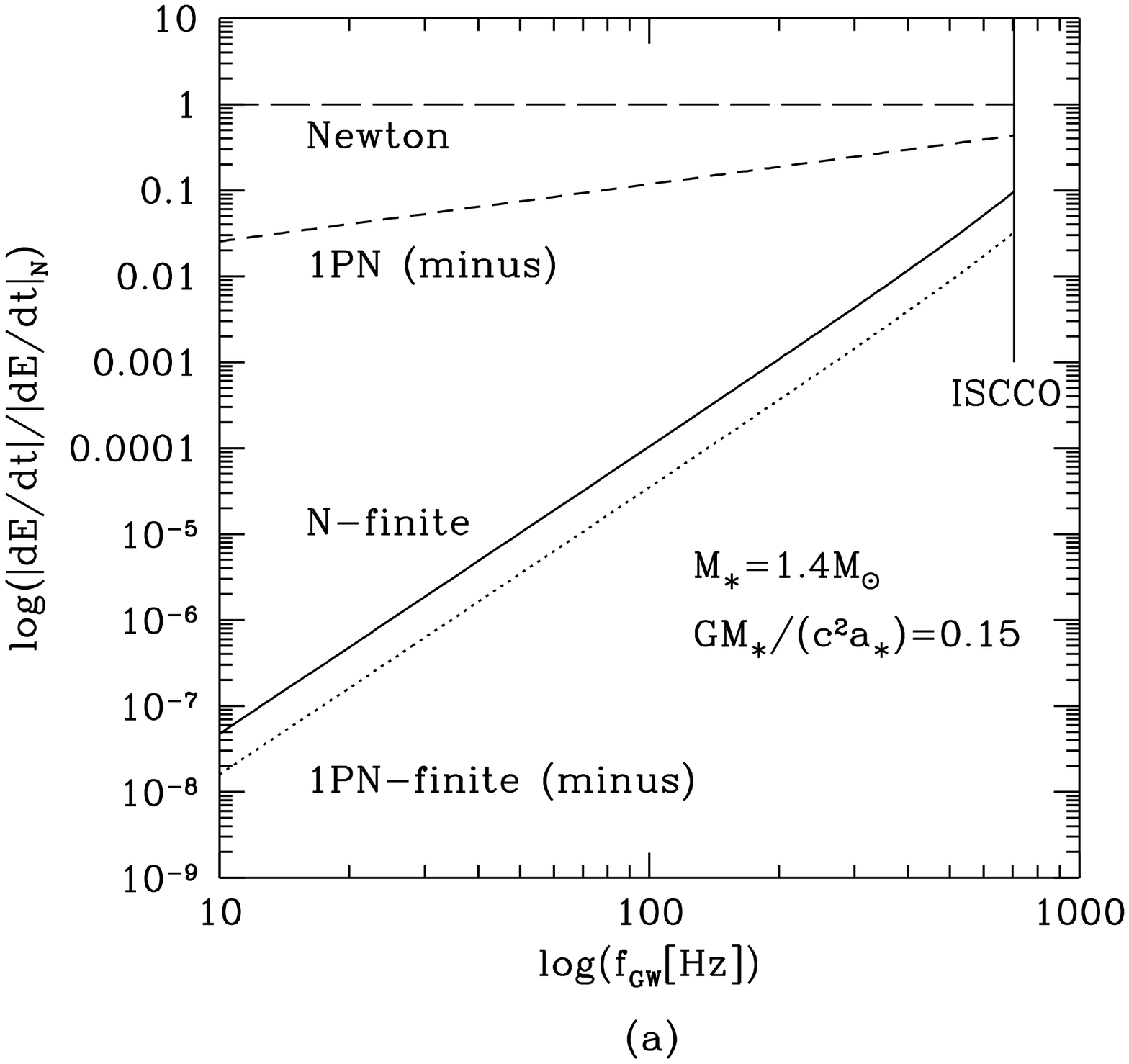}}
  \vspace{0.5cm}
  \label{1a}
\end{figure}%

\vspace{3cm}

\hspace{250pt}
Fig.1(a)

\newpage

\begin{figure}[ht]
  \vspace{1cm}
  \centerline{\epsfysize 15cm \epsfxsize 15cm \epsfbox{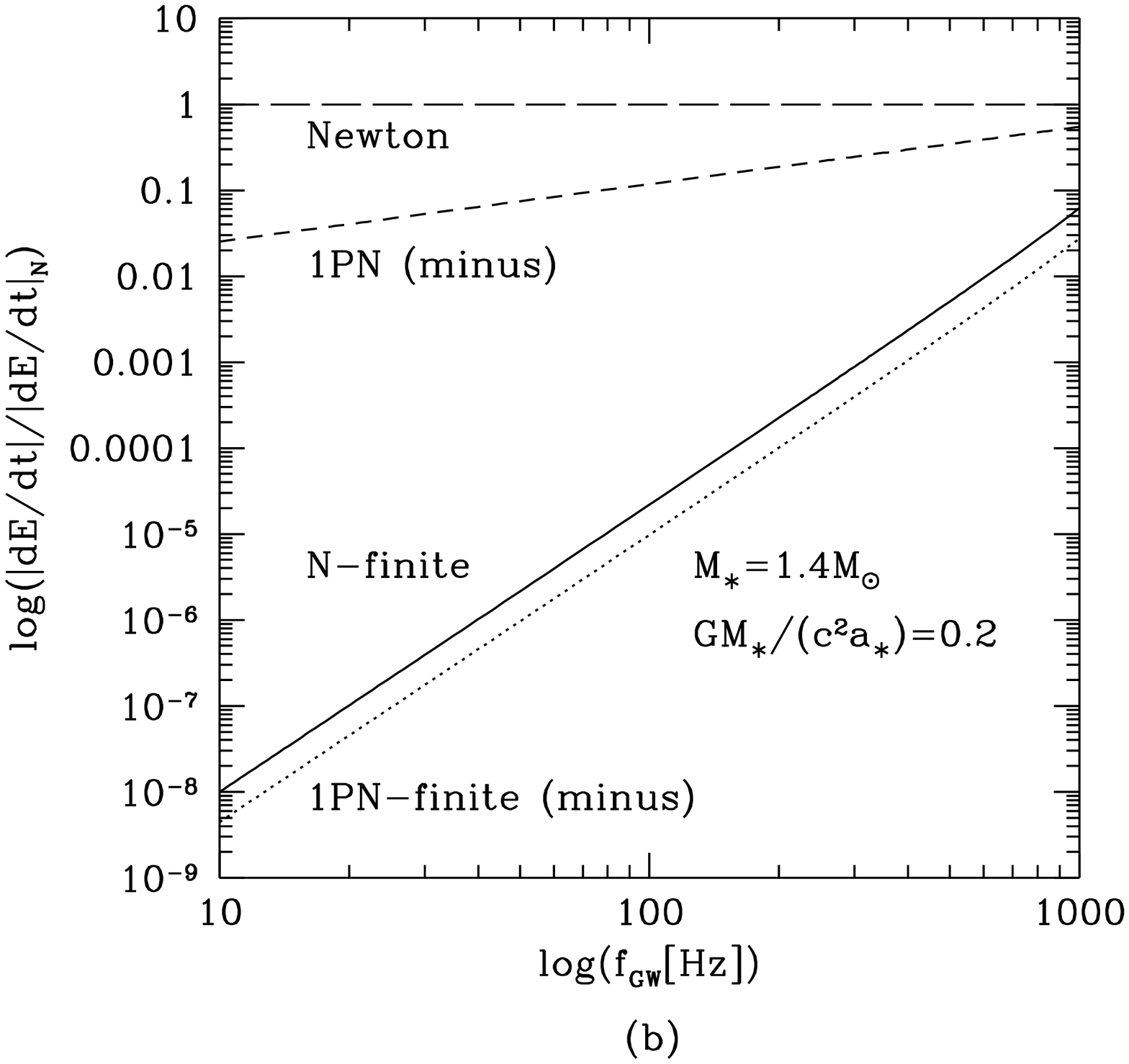}}
  \vspace{0.5cm}
  \label{1b}
\end{figure}%

\vspace{3cm}

\hspace{250pt}
Fig.1(b)

\newpage

\begin{figure}[ht]
  \vspace{1cm}
  \centerline{\epsfysize 15cm \epsfxsize 15cm \epsfbox{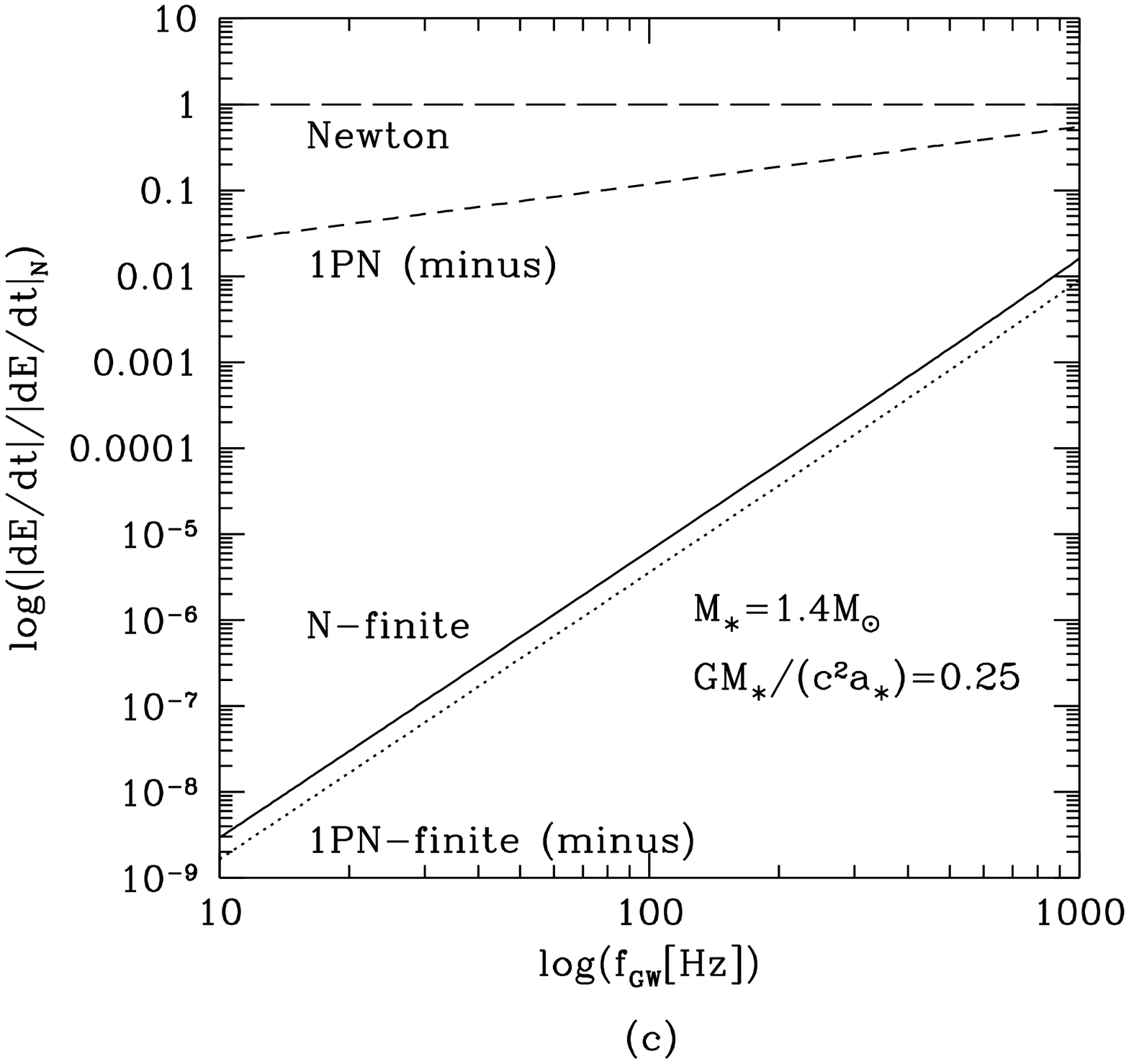}}
  \vspace{0.5cm}
  \label{1c}
\end{figure}%

\vspace{3cm}

\hspace{250pt}
Fig.1(c)

\newpage

\begin{figure}[ht]
  \vspace{1cm}
  \centerline{\epsfysize 15cm \epsfxsize 15cm \epsfbox{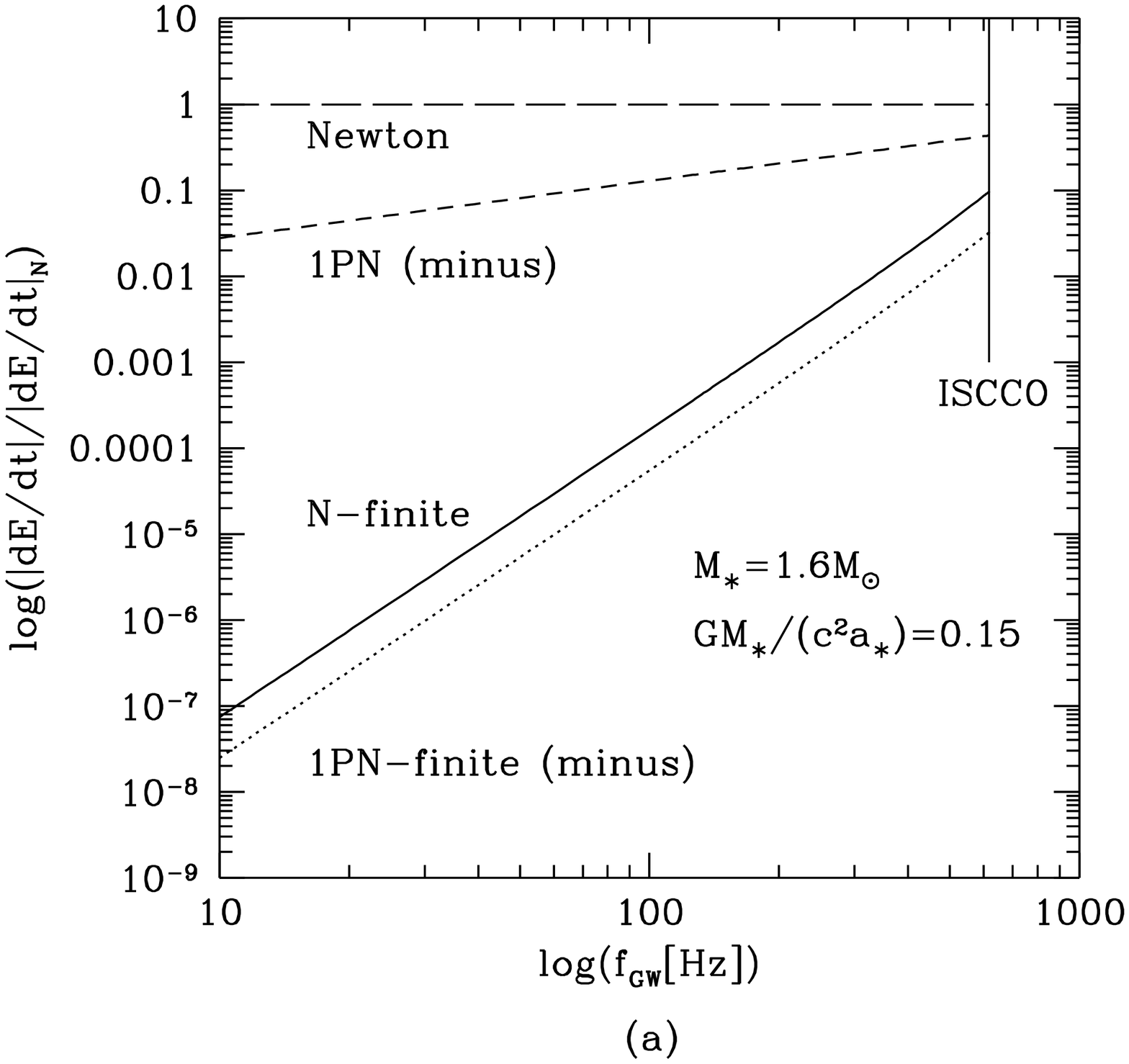}}
  \vspace{0.5cm}
  \label{2a}
\end{figure}%

\vspace{3cm}

\hspace{250pt}
Fig.2(a)

\newpage

\begin{figure}[ht]
  \vspace{1cm}
  \centerline{\epsfysize 15cm \epsfxsize 15cm \epsfbox{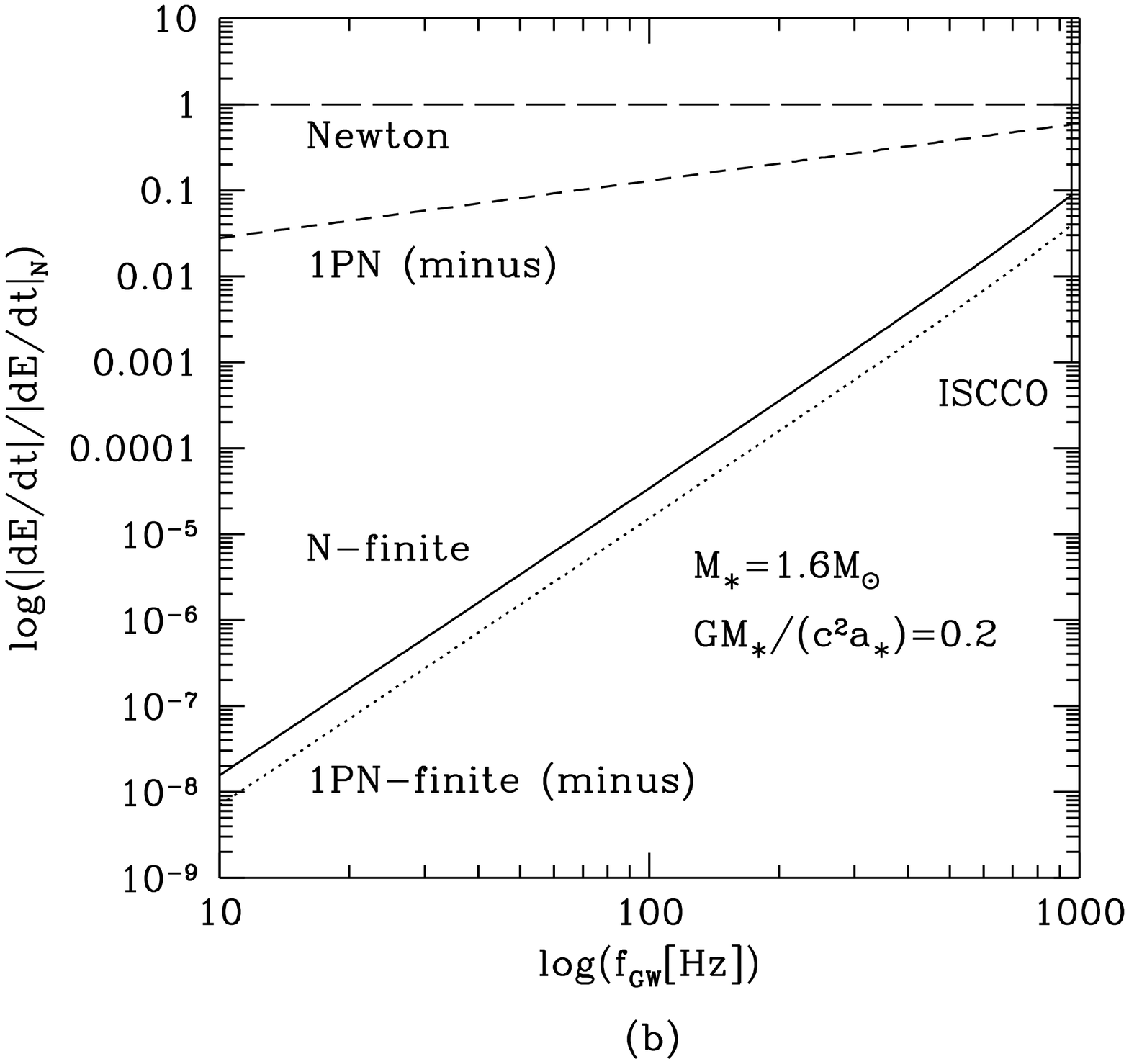}}
  \vspace{0.5cm}
  \label{2b}
\end{figure}%

\vspace{3cm}

\hspace{250pt}
Fig.2(b)

\newpage

\begin{figure}[ht]
  \vspace{1cm}
  \centerline{\epsfysize 15cm \epsfxsize 15cm \epsfbox{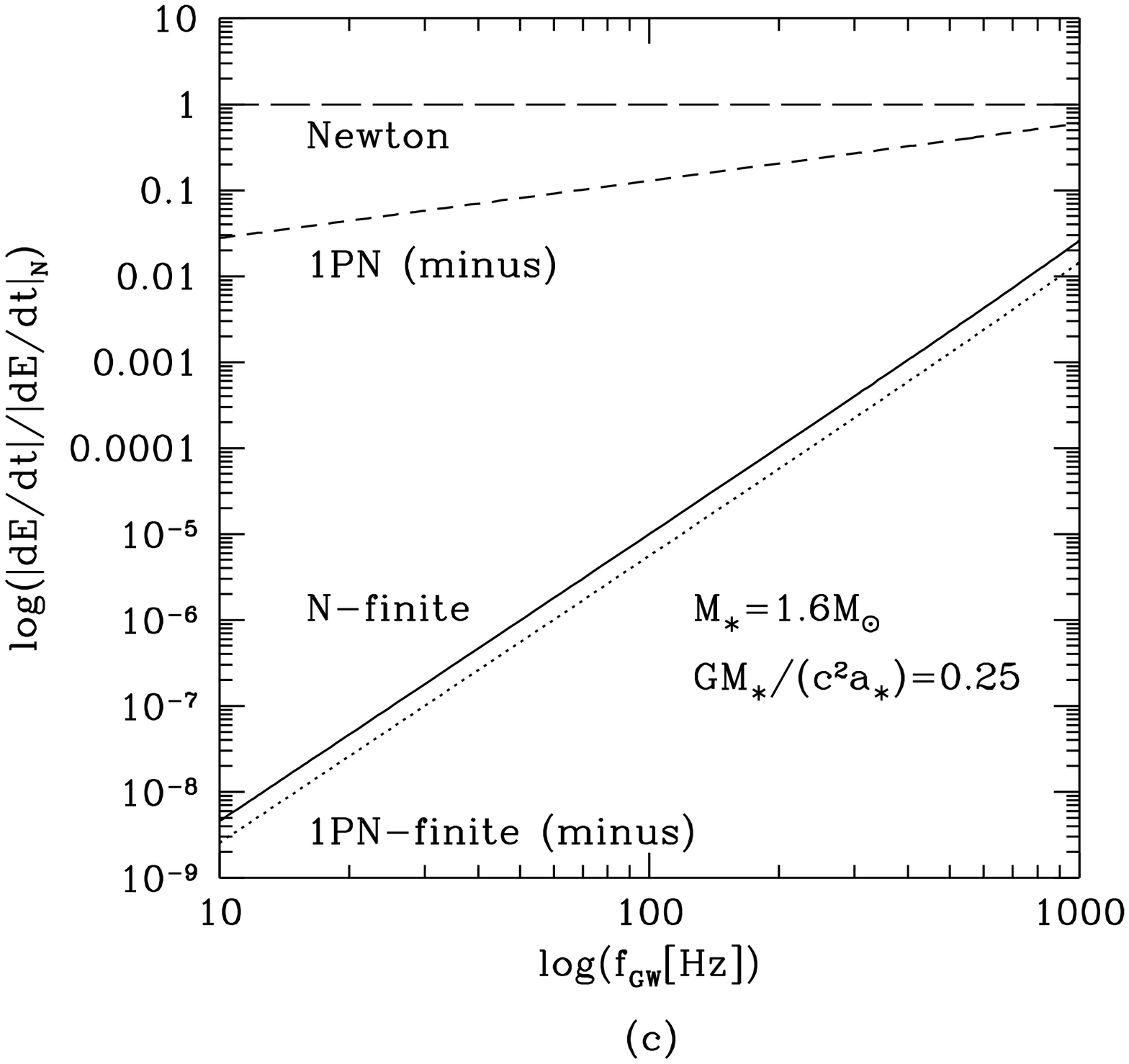}}
  \vspace{0.5cm}
  \label{2c}
\end{figure}%

\vspace{3cm}

\hspace{250pt}
Fig.2(c)

\newpage

\begin{figure}[ht]
  \vspace{1cm}
  \centerline{\epsfysize 15cm \epsfxsize 15cm \epsfbox{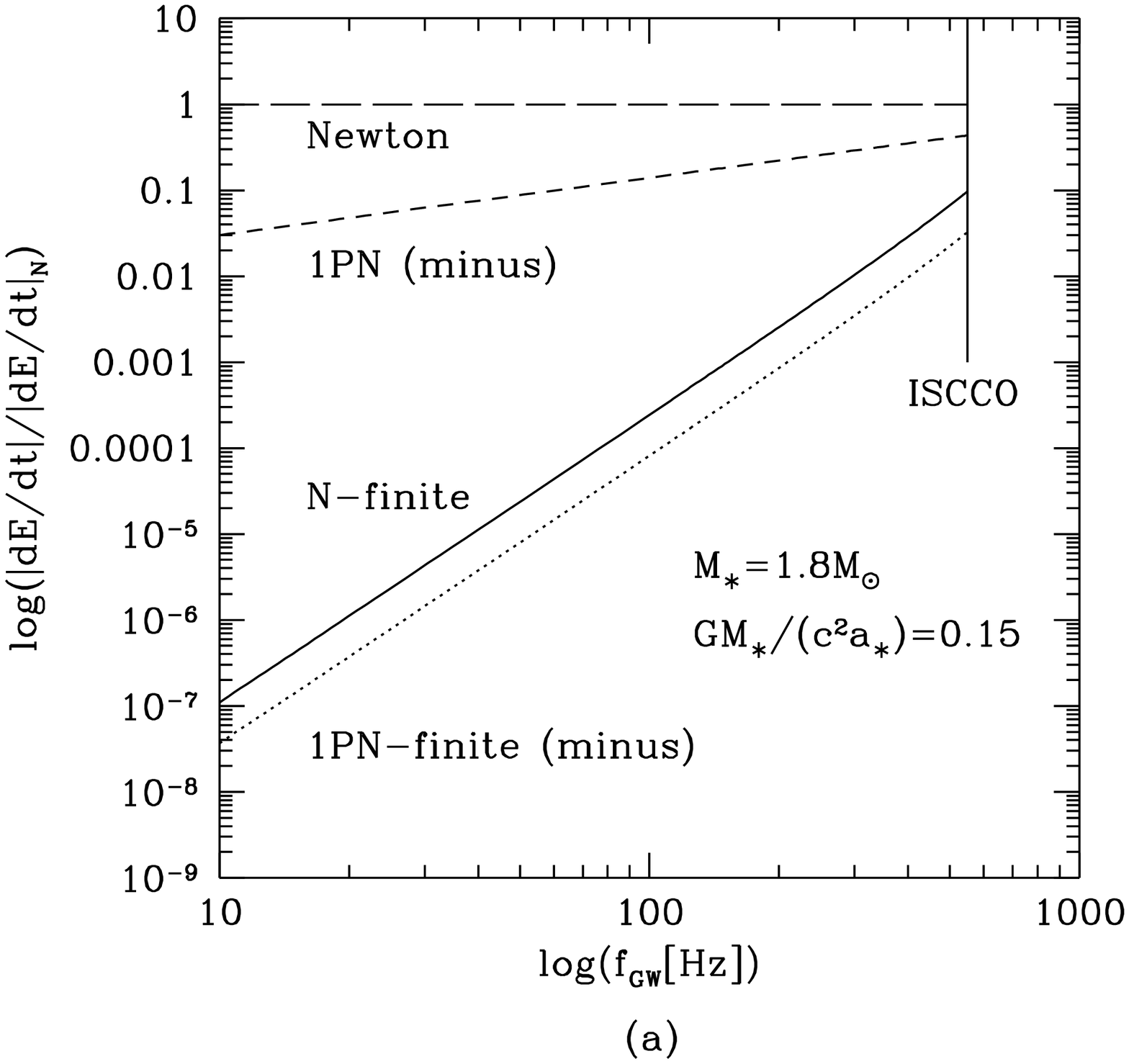}}
  \vspace{0.5cm}
  \label{3a}
\end{figure}%

\vspace{3cm}

\hspace{250pt}
Fig.3(a)

\newpage

\begin{figure}[ht]
  \vspace{1cm}
  \centerline{\epsfysize 15cm \epsfxsize 15cm \epsfbox{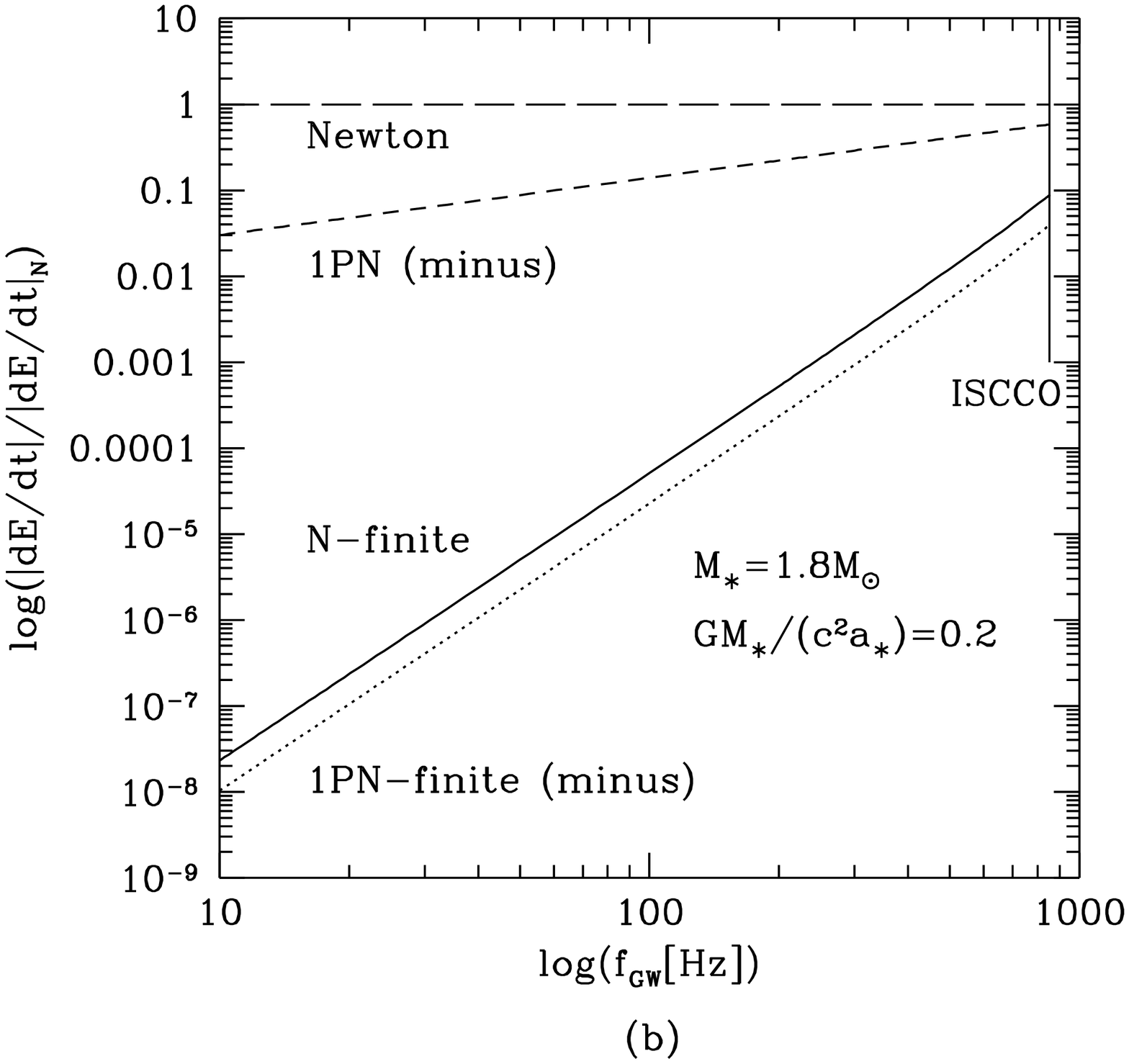}}
  \vspace{0.5cm}
  \label{3b}
\end{figure}%

\vspace{3cm}

\hspace{250pt}
Fig.3(b)

\newpage

\begin{figure}[ht]
  \vspace{1cm}
  \centerline{\epsfysize 15cm \epsfxsize 15cm \epsfbox{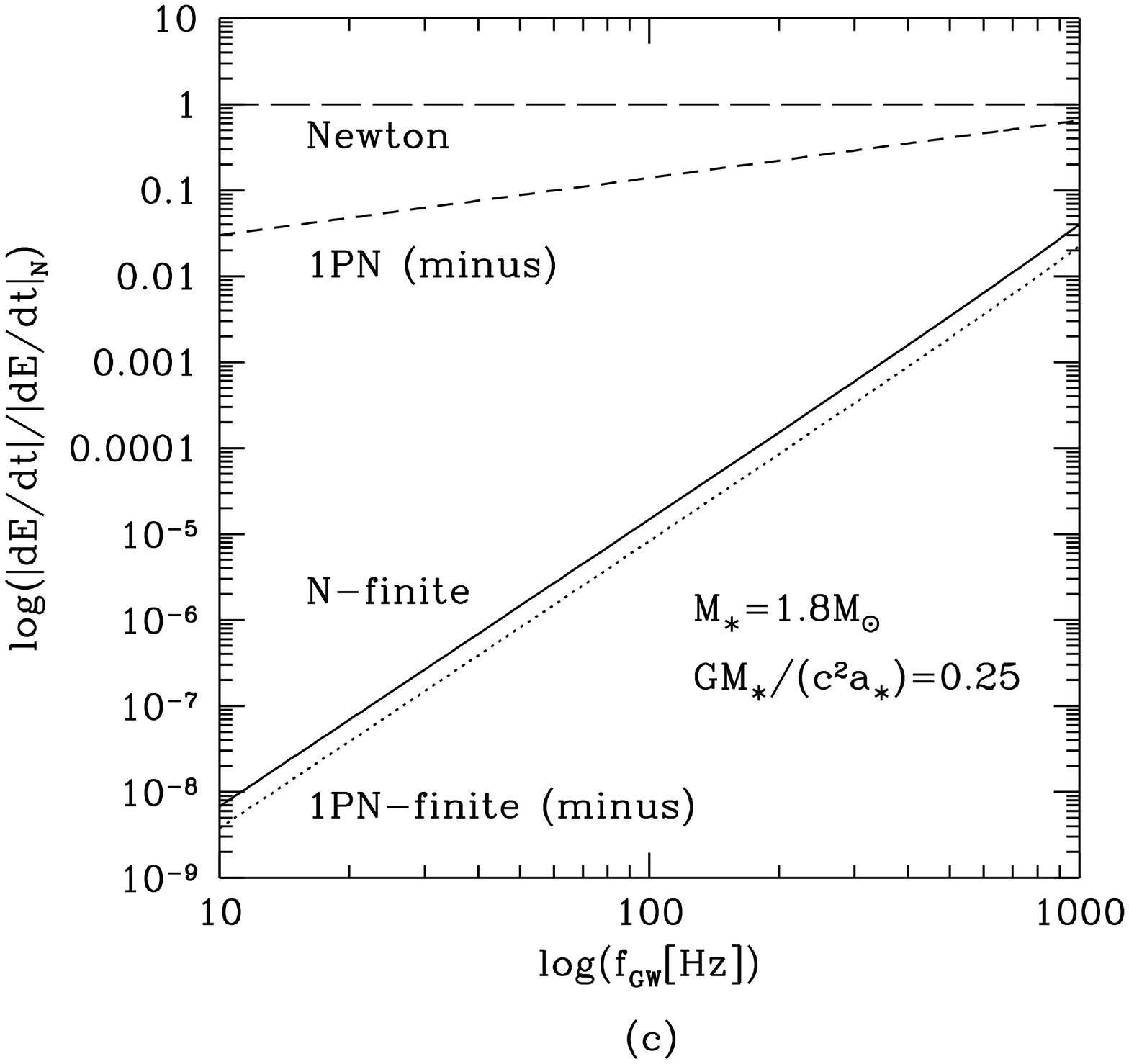}}
  \vspace{0.5cm}
  \label{3c}
\end{figure}%

\vspace{3cm}

\hspace{250pt}
Fig.3(c)

\end{document}